\documentclass[sigconf]{acmart}
\usepackage[absolute,overlay]{textpos}
\usepackage[utf8]{inputenc}
\usepackage[english]{babel}
\usepackage{booktabs} 
\usepackage{etoolbox}
\usepackage[figurename=Fig.,font={small},labelfont={bf,up}]{caption}
\usepackage[linesnumbered,ruled,vlined]{algorithm2e}
\usepackage{color, colortbl}
\usepackage{longtable}
\usepackage{xcolor}
\usepackage{soul}
\usepackage{comment}
\usepackage{url}
\usepackage{balance}
\usepackage{graphicx}
\usepackage{subfigure}
\usepackage{multirow}
\usepackage{textcomp}
\usepackage{lipsum}
\usepackage{tikz}
\usepackage{hyperref}
\usepackage{cleveref}
\usepackage{float}

\newbool{showComments}
\booltrue{showComments}

\ifbool{showComments}{%
\newcommand{\cm}[1]{\sethlcolor{yellow}\hl{[Cecilia: #1]}}
\newcommand{\af}[1]{\sethlcolor{orange}\hl{[Andrea: #1]}} 
\newcommand{\dm}[1]{\sethlcolor{lime}\hl{[Dong: #1]}}
\newcommand{\rh}[1]{\sethlcolor{cyan}\hl{[Rob: #1]}}
\newcommand{\del}[1]{\textcolor{red}{#1}}
}{
\newcommand{\cm}[1]{}
\newcommand{\dm}[1]{}
\newcommand{\af}[1]{}
\newcommand{\del}[1]{}
\newcommand{\rh}[1]{}
}

\title{\textcolor{black}{EarGate: Gait-based User Identification with In-ear Microphones}}

\newcommand{\SysName}{EarGate }

\copyrightyear{2022} 
\acmYear{2022} 
\setcopyright{rightsretained} 
\acmConference[ACM MobiCom '21]{The 27th Annual International Conference on Mobile Computing and Networking}{January 31-February 4, 2022}{New Orleans, LA, USA}
\acmBooktitle{The 27th Annual International Conference on Mobile Computing and Networking (ACM MobiCom '21), January 31-February 4, 2022, New Orleans, LA, USA}
\acmDOI{10.1145/3447993.3483240}
\acmISBN{978-1-4503-8342-4/22/01}


\begin{document}
\fancyhead{}


\settopmatter{authorsperrow=4}

\author{Andrea Ferlini}
\authornotemark[2]
\affiliation{\institution{University of Cambridge}}
\email{af679@cl.cam.ac.uk}

\author{Dong Ma}
\authornotemark[2]
\affiliation{\institution{University of Cambridge}}
\email{dm878@cl.cam.ac.uk}

\author{Robert Harle}
\affiliation{\institution{University of Cambridge}}
\email{rkh23@cl.cam.ac.uk}

\author{Cecilia Mascolo}
\affiliation{\institution{University of Cambridge}}
\email{cm542@cl.cam.ac.uk}

\author[]{}
\affiliation{\vspace{0.6in}}
\begin{textblock*}{15cm}(6.4cm,5.9cm)
$\dagger$  First authors with equal contribution in alphabetical order.
\end{textblock*}

\begin{abstract}
Human gait is a widely used biometric trait for user \textcolor{black}{identification and recognition}.
Given the wide-spreading, steady diffusion of ear-worn wearables (\textit{Earables)} as the new frontier of wearable devices, we investigate the feasibility of earable-based gait \textcolor{black}{identification}.
Specifically, we look at gait-based \textcolor{black}{identification} from the sounds induced by walking and propagated through the musculoskeletal system in the body.
Our system, EarGate, leverages an in-ear facing microphone which exploits the earable's \textit{occlusion effect} to reliably detect the user's gait from inside the ear canal, without impairing the general usage of earphones. 
With data collected from 31 subjects, we show that EarGate achieves up to 97.26\% Balanced Accuracy (BAC) with very low False Acceptance Rate (FAR) and False Rejection Rate (FRR) of 3.23\% and 2.25\%, respectively. Further, our measurement of power consumption and latency investigates how this gait \textcolor{black}{identification} model could live both as a stand-alone or cloud-coupled earable system.
\end{abstract}

\maketitle

\section{Introduction}

Human gait has been shown to be unique across individuals and hard to mimic~\cite{marsico2019survey, wan2018survey}. 
As such, there have been a variety of attempts to use gait as a biometric for user recognition and \textcolor{black}{identification}.
While computer vision based gait bio-metrics is widely spread, wearable-based gait tracking is particularly attractive for continuous \textcolor{black}{identification and, potentially, authentication}.
Wearable gait based \textcolor{black}{identification} is an enabler for various applications, including: mobile devices to remain unlocked and ready for interaction when on the owner’s person; health insurers to identify the device owner as the policy owner when rewarding healthy habits sensed via the wearable; automated entry systems for home, work or vehicles; automated ticket payment/validation for public transport~\cite{axente2020gait,johnston2015smartwatch}; etc. 
There have even been examples of using wearable gait data to generate a secure key to pair devices worn on the same body~\cite{lin2020kehkey}.

Wearable-based gait tracking methods leverage sensor data collected from wearable devices worn by the user to capture their motion dynamics, typically through accelerometer analysis~\cite{gafurov2006biometric}, or step sounds~\cite{geiger2014acoustic,altaf2015acoustic}. 
To date, the focus has been on smartphones or smartwatches as the current mass-market wearables. 
In this work, instead, we look at the use of ear-based sensing (via so-called \emph{earables}) for this task. 
The importance of the approach we present is further highlighted by the fact that, in the near-future, earables are likely to become stand-alone devices~\cite{ferlini2021enabling}. 
Hence, the need for earable-based \textcolor{black}{identification} schemes becomes more and more crucial~\cite{choudhury2021earable}.
Being able to seamlessly identify the earable wearer, it can act as an authentication accelerator for earables or mobile devices, bypassing traditional bio-metrics such as fingerprint (requires the integration of capacitive sensing pad on earables with limited size) and face recognition (impossible to capture front face image from an earbud). Occasionally, if the identity is mistakenly rejected, a request for a secondary authentication can be triggered. The secondary authentication method could be implemented on a companion device, such as smartphone/smartwatch, that has full access to fingerprint and human face.
Additionally, once the user has been successfully identified by the earable, the earable itself can act as a hub to authenticate the user for access control (e.g. opening their office door, validating their ticket at the train station, etc.).
Furthermore, with the increased sensing capabilities of earables, successfully identifying the earable's wearer becomes crucial in order to associate the sensitive bio-medical information collected by the earable to the right user.

Today's earables fall broadly into two categories: advanced playback devices (e.g., Apple AirPods) or assistive devices (e.g., advanced hearing aids).
Notably, although being useful for all earables, such continuous \textcolor{black}{identification} systems suit hearing aids particularly well, as they are continuously worn throughout the day.
Earables present potential advantages over other wearables.
For instance, earables' location in-the-ear means that, unlike phones, they are in a consistent, stable location, firmly attached to the user’s body; and unlike watches, they are not subject to confounding movements during walking (e.g., carrying objects, using tools or loose straps). 
Further, they also offer two independent signals (one per each ear). 

Here we investigate acoustic-gait as a convenient alternative to inertial-based gait tracking.
Specifically, we look at the possibility of gait-based \textcolor{black}{identification} from the \emph{sounds} caused by the physical act of walking and transmitted internally via the musculoskeletal system. 
Our novel earable-based \textit{acoustic-gait} \textcolor{black}{identification} system, EarGate, is built around a cheap in-ear facing microphone that is already available on most earbuds and hearing aids (e.g., for noise cancellation purposes).
We show that an earable equipped with a microphone \emph{inside} the ear canal can not only detect motion signals, but those are also amplified by the combination of two biological phenomena: \textit{bone conduction} and the \textit{occlusion effect}. 
We conducted a user study, with 31 subjects of mixed gender, demonstrating how the acoustic-gait thus collected is a good candidate for a privacy-preserving \textcolor{black}{identification} system (benefit from the in-ear facing microphone). 
Further, we investigated the implications of running the \textcolor{black}{identification}-framework entirely on-device, as well as offloading the computation (either to the Cloud or to a companion device) to show the versatility of the approach.

In summary, {\em our approach leverages sensors (microphones) already inherently  embedded in earables for main functions}. As mentioned, previous gait \textcolor{black}{identification}/recognition approaches have often relied on inertial sensors~\cite{gafurov2006biometric}. 
However, inertial sensors have made reasonable penetration into the high-end leisure devices market but not as much into cheaper earables, or the hearing aid market: the addition of inertial sensors to these devices would entail more complex system design and form factor~\cite{zhang2019motion}, as a consequence, inertial-based earable solutions are likely to result in increased cost and delays in reaching the market. 
This paper offers an investigation into an alternative to inertial sensors through the use of microphones.
Notably, while we acknowledge the merits of inertial-based gait recognition approaches, there is great value in showing the potential of a lesser explored modality, such as in-ear microphones.
Particularly given what suggested by research and market trends, according to which, in the near future, miniaturized form factor will have a key role, especially as the distinction between hearing aids and earables is likely to become less marked. 
Further, in-ear based microphones offer complementary advantages as an overall lower cost, given its importance as a sensor to enable noise cancellation, a must-have feature for both high-end leisure earables and in hearing aids.
While in this paper we focus on acoustic-gait recognition, we note this is only one of the possible use cases for in-ear bone-conducted sounds.
For instance, these could indeed be used for activity recognition~\cite{ma2021oesense}, as well as physiological sensing.
In particular, the contributions of this work can be summarized as:
\begin{itemize}
        \item We devise a novel earable-based gait \textcolor{black}{identification} system, EarGate, consisting of  a hardware prototype and a software pipeline. \textcolor{black}
        EarGate, not only leverages a novel type of signal, in-ear bone-conducted sounds, to identify users based on their gait, but also is accurate, robust, and has improved usability (only a few steps are sufficient to identify the user, without the need for them to be continuously walking);
        \item To track gait cycles from earables, we leverage the occlusion effect, a natural enhancement of low-frequency components in an occluded ear canal~\cite{stone2014technique}. In addition, we designed an end-to-end signal processing pipeline and some techniques to guarantee the reliable presence of the occlusion effect;
        \item We collected a one-of-its-kind dataset with 31 subjects under various conditions, which we released to the research community at Kaggle\footnote{https://www.kaggle.com/dongma878/eargate};
        \item We evaluate the \textcolor{black}{identification} performance of EarGate under various practical scenarios, showing we can achieve up to 97.26\% Balanced Accuracy (BAC) with very low False Acceptance Rate (FAR) and False Rejection Rate (FRR) of 3.23\% and 2.25\%, respectively. Furthermore, we demonstrated that EarGate is robust to high-frequency internal (human speech) and external (music playback and phone calls) noises;
        \item Finally, we assessed the system performance of EarGate by measuring the power consumption and latency. We find EarGate can work in real-time (74.25~ms on-device \textcolor{black}{identification} latency) consuming acceptable energy (167.27mJ for one-time on-device \textcolor{black}{identification}). %
        This confirms that EarGate could be deployed in potential new generation earables which will likely be standalone from the system perspective.
    \end{itemize}
\section{Preliminaries}
\label{sec:preliminary}
In this section, we first brief the reader on the rationale driving our work and then provide evidence of the feasibility of our approach.

\subsection{Rationale: Occlusion Effect}
\label{sec:occlusion-effect}
In this work, we leverage the natural low-frequency boost (up to 40~dB depending on the frequency~\cite{carillo2020theoretical}) provided by the phenomenon known by the name of \textit{occlusion effect}. This section provides the reader with a basic understanding of what the occlusion effect is in practice and how it can be exploited to facilitate in-ear human gait \textcolor{black}{identification}. 

\begin{figure}[]
	\centering
	\includegraphics[scale = 0.23]{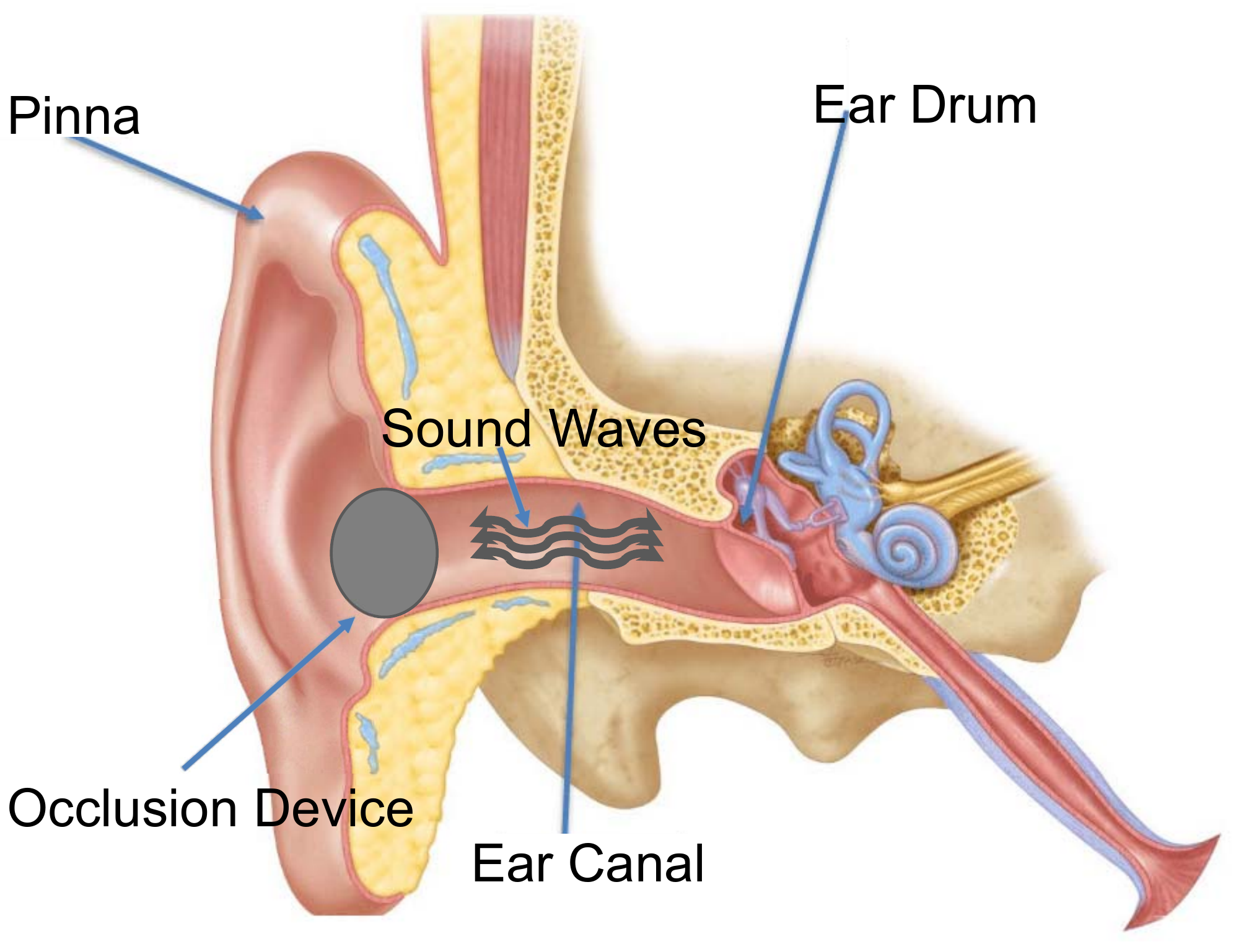}
	\vspace{-0.15in}
	\caption{Anatomy of human ear and occlusion effect. When the orifice is occluded, sounds are trapped in the ear canal, resulting in the amplification of low frequency components.}
	\label{fig:occlusion_effect}
	\vspace{-0.15in}
\end{figure}

From a physiological point of view, the occlusion effect can be defined as a dominance of the low-frequency components of a bone-propagated sound due to the loss of relevance of the outer ear sound pathways whenever the ear canal orifice is sealed (i.e.~occluded)~\cite{stone2014technique}. For example, such phenomenon could be experienced in the form of echo-like/booming sounds of their own voice if a person is speaking and is obstructing her ear canals with a finger or an earplug. Essentially, sound is nothing but vibrations propagating as acoustic waves. Usually, it travels through bones and escapes the inner-ear via the ear canal orifice. However, if this opening is obstructed, the vibration waves are blocked inside the canal and are bounced back to the eardrum~\cite{stenfelt2011acoustic}, as illustrated in Figure~\ref{fig:occlusion_effect}. As a result of that, the low-frequency-bone-conducted sounds are amplified~\cite{schlieper2019relationship}. A more precise definition of what the occlusion effect entails can be denoted by the ratio between the sound pressure inside the occluded ear canal and that in the open ear~\cite{stenfelt2007model}. Specifically to our use case, the vibrations generated by a foot hitting the ground, as soon as a person takes a step (basic component of a gait cycle), propagate through the body via bone-conduction. 
Interestingly, these vibrations are amplified if the ear canal of the person is occluded by, for example, an earbud.

\subsection{Out-ear Mic vs. In-ear Mic}
Compared to the approaches using external microphones (out-ear mic) for gait recognition~\cite{geiger2014acoustic,altaf2015acoustic}, the use of occlusion effect and an inward-facing microphone brings the following advantages: \textbf{(1)} given an occluded ear canal, an inward-facing microphone, which mostly records bone-conducted sounds, results to be less susceptible to external sound and consequent environmental noise. This not only means our system is more robust to noise, but practically, it makes our approach more appealing from a privacy perspective: potentially sensitive external sounds, such as human speech, are hardly audible from our in-ear facing microphone.
\textbf{(2)} another direct consequence of the occluded ear canal, is that the body-sounds we are interested in are relatively low-frequency and, therefore, we greatly benefit from the low-frequency amplification boost induced by the occlusion effect, resulting in an improved signal-to-noise-ratio (SNR) of the desired signal. 

\begin{figure}[]
	\centering
    \includegraphics[width=.99\linewidth]{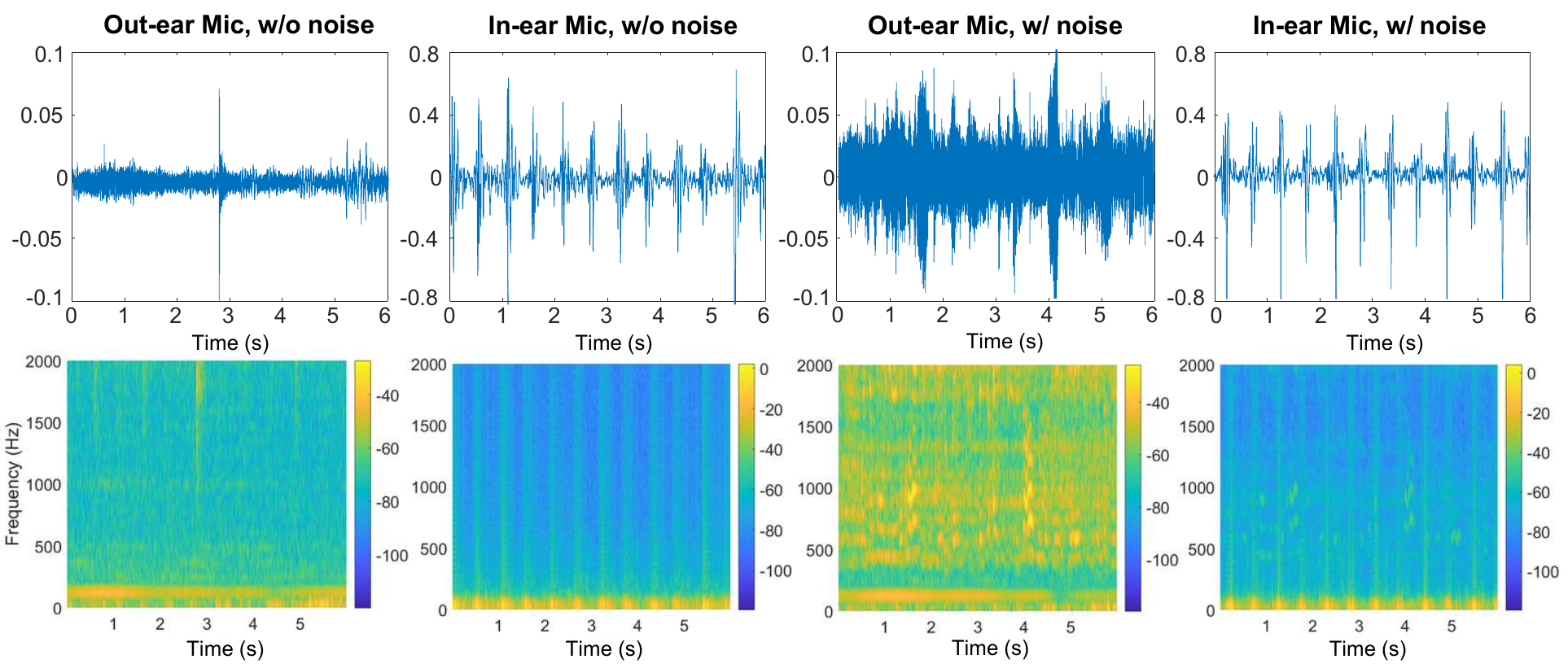}
    \vspace{-0.1in}
    \caption{Walking signals collected with out-ear and in-ear microphones under different conditions. The in-ear microphone data show higher gain at low frequencies (<50~Hz) due to the occlusion effect, and more resilience towards environmental noise.}
	\label{fig:raw_compare}
\end{figure}

\Cref{fig:raw_compare} plots the raw signals and corresponding spectrograms collected with the out-ear microphone and in-ear microphone when a subject is walking, with and without environmental noise.
From these graphs, we can clearly observe how the in-ear microphone overcomes the drawbacks of the out-ear microphone that captures air-conducted sounds. First, air conduction incurs large attenuation, while bone conduction and occlusion effect guarantee an excellent SNR. Second, the walking sound measured by the out-ear microphone resides in higher frequency (audible) and is completely mixed with other environmental noise (e.g., music and human speech). Consequently, they can not be separated even with a lowpass filter.

\begin{figure}[]
	\centering
	\subfigure[]{
	    \centering
	    \includegraphics[scale = 0.22]{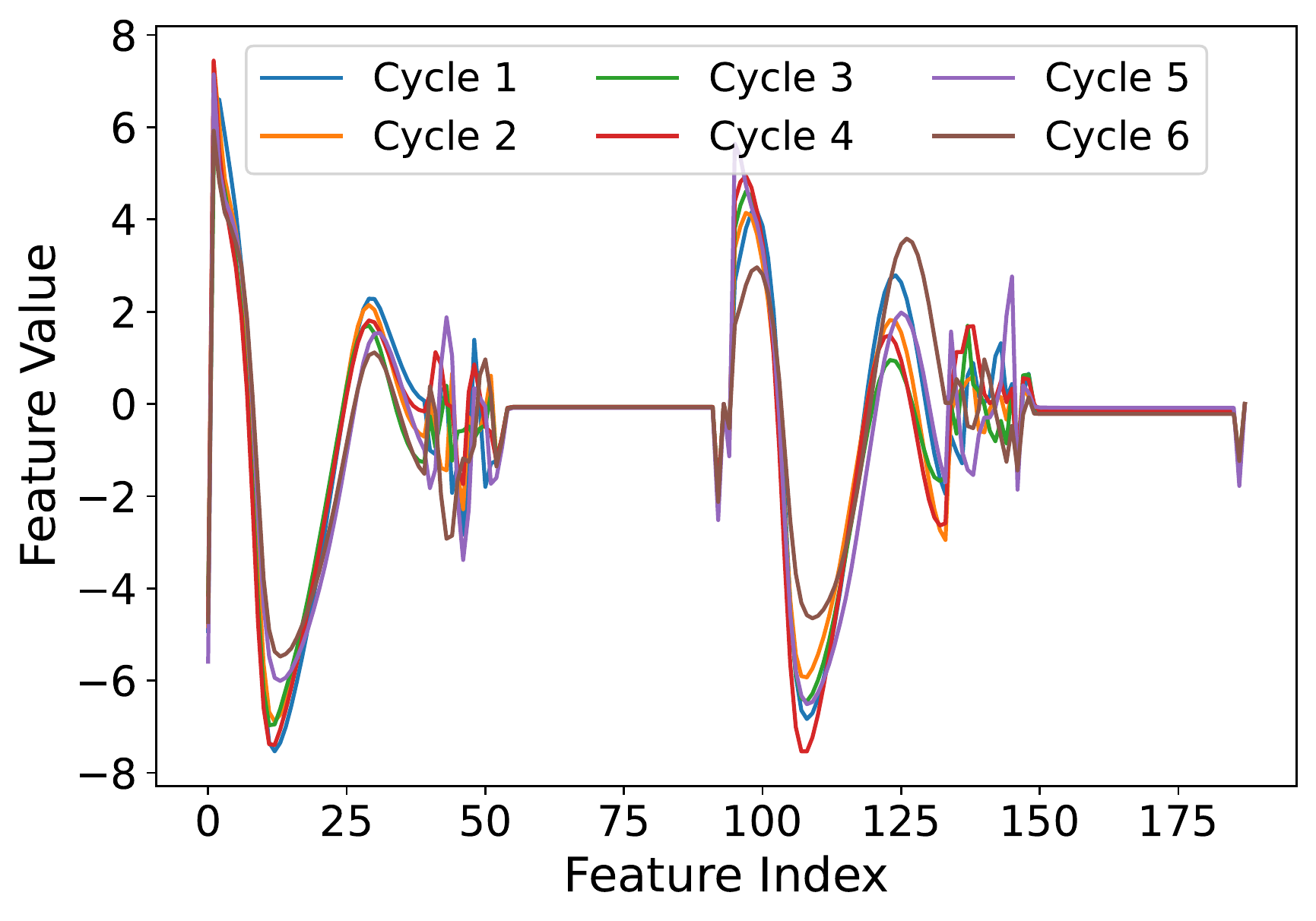}
	    \label{fig:feasibility_same_subject}}
    \subfigure[]{
	    \centering
	    \includegraphics[scale = 0.22]{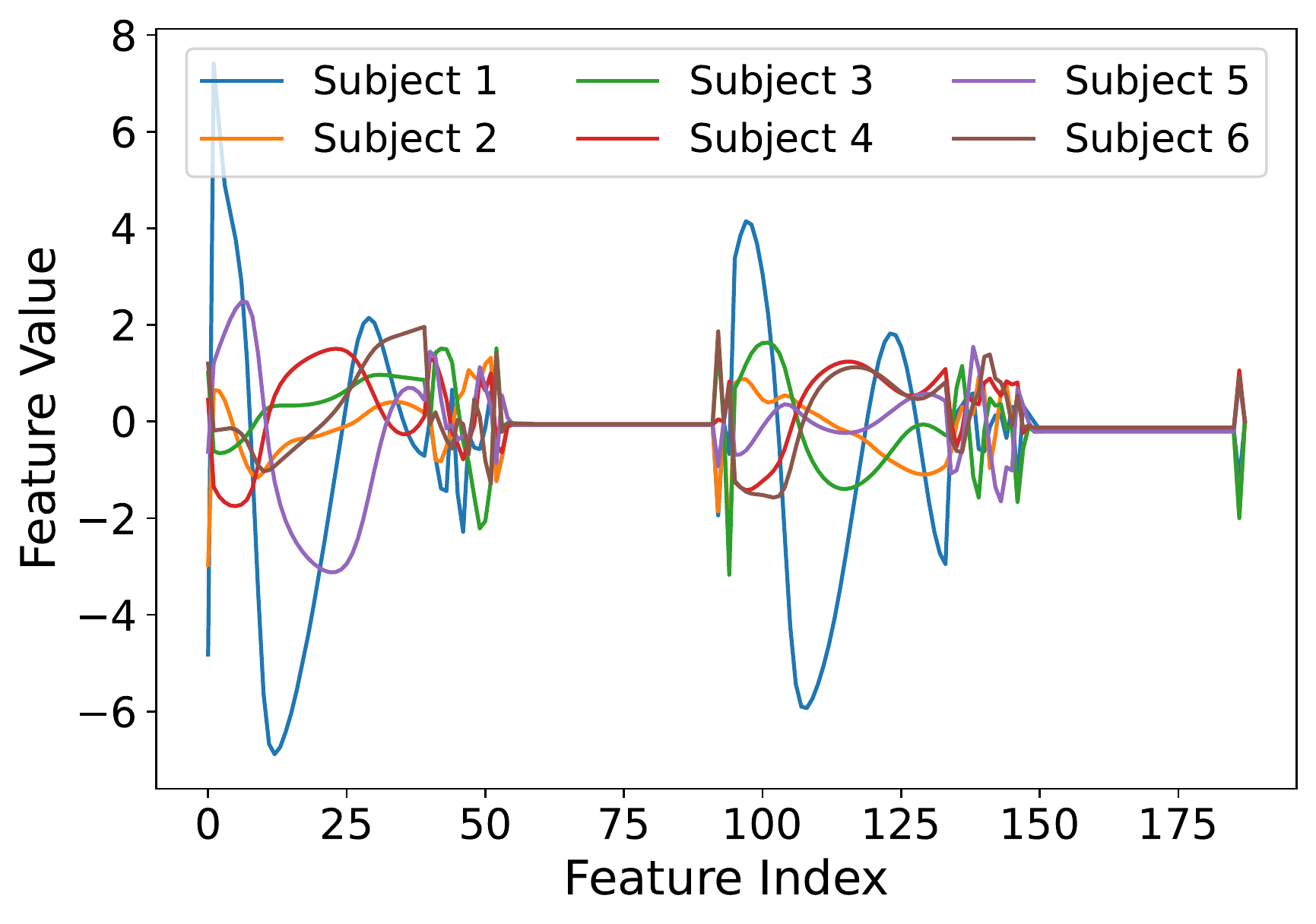}
	    \label{fig:feasibility_four_subject}}
	\caption{
	(a) Extracted feature vectors for four different gait cycles from the same subject, and (b) gait cycles from four different subjects. \Cref{fig:feasibility_same_subject} clearly shows how the gait cycles captured by \SysName are consistent for each individual (i.e.~high intra-class similarity). \Cref{fig:feasibility_four_subject}, on the other hand, shows how the gait cycles are distinguishable among subjects (i.e.~high inter-class difference).}
	 \label{fig:feasibility}
\end{figure}

\subsection{\textcolor{black}{Human Gait Primer}}
\label{sec:gait-primer}
The walking style of a person is commonly known as their gait.
Medical and physiological studies~\cite{wan2018survey} suggest that the human gait shows 24 different components. 
The differences between the gait of distinct subjects are caused by the uniqueness in their muscular-skeletal structure.
The human gait is regulated by precise bio-physical rules~\cite{marsico2019survey}. 
These, in turn, are dictated by the tension generated by the muscle activation and the consequent movement of the joints.
As a result of that, the forces and moments linked to the movement of the joints cause the movement of the skeletal links which, therefore, exert forces on the environment (e.g. the foot striking the ground).
Hence, the human gait can be described as a generation of ground-reaction forces which are strongly correlated with the muscular-skeletal structure of each individual.
In practice, differences in the body structure of individuals are among the factors that produce the interpersonal differences in walking patterns that enable gait-based identification.

\subsection{Feasibility Exploration}
\label{sec:feasibility}
Although the in-ear microphone is capable of detecting human steps, whether we could extract the acoustic gait to differentiate people remains unclear. To demonstrate the feasibility of gait recognition, we need to prove that (1) gait cycles belonging to the same individual are consistent with each other (i.e., intra-class similarity) and (2) gait cycles belonging to different subjects show significantly different patterns (i.e., inter-class dissimilarity). Thus, we collected data from four subjects and extract features (see \Cref{sec:features}) to represent the user's gait. As shown in \Cref{fig:feasibility}, the extracted features exhibit high intra-class similarity and high inter-class difference. Therefore, it would be feasible to identify people with the acoustic signals measured with the in-ear microphone.
\section{System Design}\label{sec:sys_design}
This section presents an overview of \SysName and its functionalities and a description of the proposed gait-based \textcolor{black}{identification} pipeline.
 
\subsection{EarGate: System at a Glance}
\begin{figure}[]
	\centering
    \includegraphics[width=.8\linewidth]{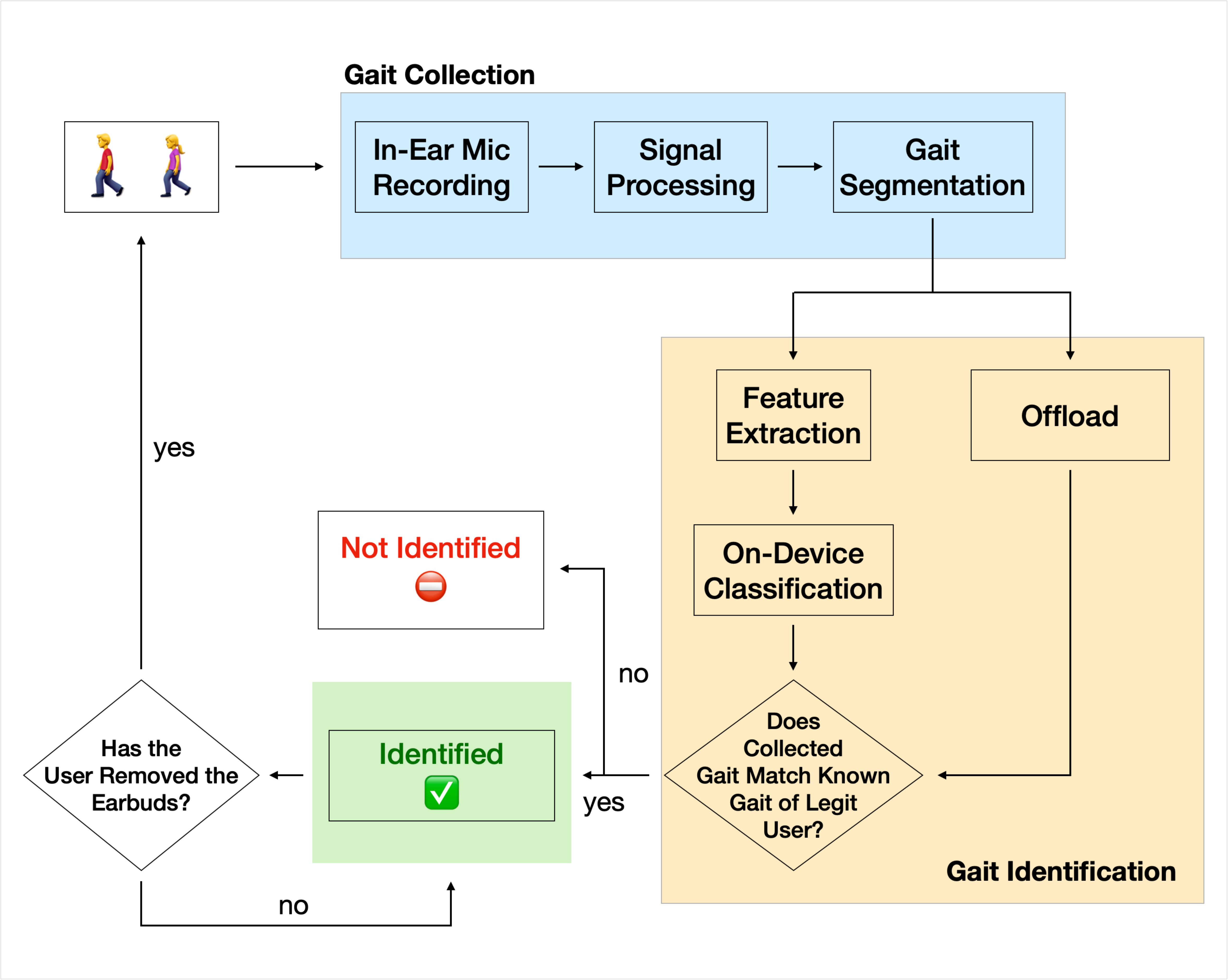}
    \vspace{-0.1in}
	\caption{EarGate Functioning.}
	\label{fig:sys_flowchart}
\end{figure}

Initially, the user has to take part in an enrollment phase, a required stage during which the system acquires the data and process them (pre-processing and feature extraction) before training the model to recognize the acoustic-gait of the legitimate user.
Notably, as we will discuss in \Cref{sec:training_size}, a small number of steps and, therefore, little time, is sufficient to achieve good \textcolor{black}{identification} performance.
Once the enrollment phase is over, the system is ready to operate.
As shown in~\Cref{fig:sys_flowchart}, EarGate silently collects, and pre-processes on-device, acoustic-gait data. 
The system is provisioned to either execute the (\textit{limited}) computation required to \textcolor{black}{identify} the user (or reject them) on-device, or to offload it, e.g., to the cloud, a smartphone, or a remote server (\Cref{sec:system_eval}).

Notably, there is NO need for the user to be \textit{constantly} and \textit{instantly} walking to be \textcolor{black}{identified}. 
Instead, \SysName can perform a one-time \textcolor{black}{identification} once the user wears the earbuds and walks a couple of steps (e.g. to grab a coffee, to go to the restroom, etc.); the \textcolor{black}{identification} result (either \textcolor{black}{recognized} as legit user, or not) is then considered valid until the user removes the earbuds (i.e., the only chance the wearer has of switching identity).
Detecting such occurrence is, in practice, very simple, as the natural properties of the occlusion effect will suddenly disappear from the in-ear microphone recordings, whenever the user removes the earbuds. 
Besides, commercially available earbuds, like Apple AirPods already do that (i.e., to automatically stop the music whenever the user removes their earbuds).
Such scheme significantly relieves the user burden of \textit{"walk now to be \textcolor{black}{identified}"} and saves system energy as one \textcolor{black}{identification} could be valid for a longer time.  

\subsection{Signal Processing}
\label{sec:signal_processing}
\begin{figure}[]
	\centering
	\includegraphics[scale = 0.30]{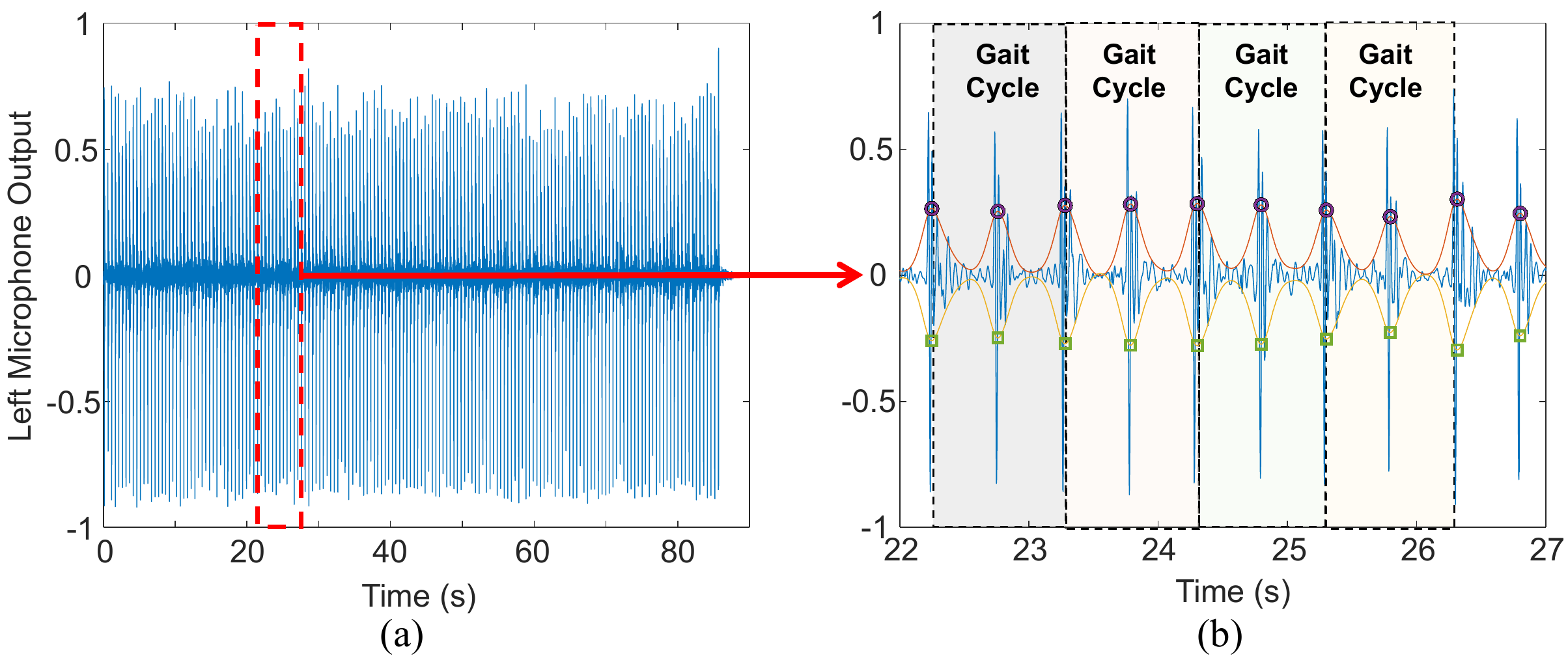}
	\vspace{-10pt}
	\caption{ (a) The low-pass filtered signal collected when one participant is walking, (b) a segment showing the performance of proposed gait segmentation algorithm.}
	\vspace{-0.15in}
	\label{fig:gait_segmentation}
\end{figure}

Our prototype records the microphone outputs in 48~kHz, while the generated step sounds are at relatively low frequencies
\footnote{Note that 'frequency' in this paper refers to the pitch of the step sound, instead of the cadence/speed of walking.}. 
To minimize the computation overhead during processing, we first down-sample the recorded data to 4~KHz. Then, a low-pass filter with a cutoff frequency of 50~Hz is applied to eliminate the high-frequency noise.
The choice of 50~Hz cutoff frequency is to guarantee a good signal to noise ratio of the walking signal, while retaining robustness to most environmental noise (typically higher than 50~Hz). Moreover, due to the large attenuation of sound in the air as well as the blockage of the ear canal opening, the majority of the noise is suppressed or canceled in the ear canal.
\Cref{fig:gait_segmentation} shows the low-pass filtered signal from the left earbud when one participant is walking on tiles with sneakers. We can observe that an acoustic gait cycle is composed of two spikes (happening when the foot hit the floor in the \textit{strike phase}) and two relatively flat (silent) periods (denoting the \textit{swing phase}). This is different from sinusoidal-like patterns recorded by the IMU~\cite{prakash2019stear}, therefore, existing gait segmentation approaches proposed for IMU data are not applicable. We propose a peak detection based algorithm to segment the signals and extract gait cycles.  

Specifically, we first use the Hilbert transform to extract the envelopes of the filtered signal and apply a low-pass filter (with a cutoff frequency of 3~Hz) on the envelops to smooth them, as illustrated by the red (upper envelop) and yellow (lower envelop) curves in \Cref{fig:gait_segmentation}~(b). Then, we perform peak detection on the filtered envelops and regard the peaks as the points when the human foot hits the ground. 
Whenever a pair of upper peak-lower peak is aligned, we treat it as a step.
Next, we select the cycle start points by skipping one between every two peaks, as each gait cycle consists of two steps. Lastly, a gait cycle is extracted as the samples between every two cycle start points. Most of the extract cycles last for around one second, so we interpolate them into the same length of 4,000 samples using spline interpolation.  

\subsection{Feature Extraction}\label{sec:features}
Then, \SysName extracts features that could represent the characteristics of user gait from each cycle. We do that using \textit{librosa}~\cite{librosa}, a Python package specific for audio processing. 
Specifically, we looked at:
\begin{itemize}
    \item Mel-Frequency Cepstral Coefficients (MFCC): obtained from
the short-term power spectrum, MFCC certainly is one of the most common  and known features in audio processing~\cite{davis1980comparison};
    \item Chroma of Short-Time Fourier Transform (STFT);
    \item Mel Spectrogram: the signal spectrogram in the Mel-scale;
    \item Root-Mean-Squared Energy (RMSE): the Root-Mean-Squared (RMS) of the STFT of the signal, which provides information about the power of the signal;
    \item Onsets: the peaks from an onset strength envelope, result of a summation of positive first-order differences of every Mel-band.
\end{itemize}

Rather than only using the data recorded by either the left-earbud-microphone or those coming from the right-earbud-microphone, separately, we fuse them concatenating the features we extracted from each.
Notably, unlike other wearables (e.g., smartwatches) earables are two and, therefore, it is possible to leverage, and fuse, their two independent measurements of the same phenomenon.

\subsection{\textcolor{black}{Identification} Methodology}
Like most \textcolor{black}{identification} systems (e.g., FaceID and TouchID), before online \textcolor{black}{identification}, EarGate requires an enrollment phase.
It is during this phase that the legitimate users provide their gait data to system, thus training a model to classify the users as either legitimate users, or impostors.
Notably, all the users that have not been seen by the model in the enrollment phase will be regarded as impostors.
In this work, we consider two enrollment schemes: \textbf{(i)} with and \textbf{(ii)} without impostor data.
The former leverages a pre-trained model with benchmark impostor data, together with some data from the legitimate user. 
However, given it is not always possible to assume the availability of benchmark imposer data, especially little after the release of the system, we also look at a model solely trained on the legitimate user data. 
In either case, the system needs some walking data from the legitimate user and, therefore, will instruct the user to follow an enrollment protocol (basically walking for a few steps and training a model with such walking data).
For the online-\textcolor{black}{identification} framework we adopt: \textbf{(i)} a two-class Support Vector Machine (SVM) classifier (if benchmark imposer data are available); or \textbf{(i)} a one-class SVM classifier (when we only have the data of the legitimate user), due to its high computational efficiency and low complexity~\cite{gao2019earecho}.

Prior work unequivocally showed how gait is a unique user fingerprint, and how is very hard for an impostor to impersonate another person's gait~\cite{mjaaland2010walk,muaaz2017smartphone}.
Hence, the aim of this paper is to use earables as a personal-\textcolor{black}{identification} device, we sought to assess whether our in-ear microphone-based approach is capable of recognizing the user in such way that, if others (i.e.~\textit{impostors}) were using the device, it would be able to spot it.
To this extent, we consider both \textbf{(i)} {\em Replace Attacks} (a different user mistakenly tries to use the earables) and \textbf{(ii)} {\em Mimic Attacks} (a malicious attacker deliberately tries to use the earbuds by actually impersonating the user, i.e., simulating the user's gait).
As highlighted by our evaluation, different users can be distinguished very clearly by our system, thus making mimic attacks even more unfeasible.
To further clarify that, for example, let us assume there is a very well-trained imposer, who can accurately mimic the gait of the legitimate user, generating the very same vibrations whenever the feet hit the ground. 
However, even given all the most favorable conditions, when the vibrations propagate through the body and the bones, all the way to the ear canal, they will inevitably be different from those belonging to the legitimate user. 
This is because, as discussed in~\Cref{sec:occlusion-effect}, the human body and skeleton act as a natural modulator.
Hence, in the remainder of the paper we will focus on showing the performance of our system against the more common replace attacks.
\section{Implementation}
\label{sec:implementation}
This section provides the design details our prototype and describes the data collection procedures we followed.
\subsection{EarGate Prototype}
To have full control on the data (sampling rate as well as unlimited access), we decided to build our own prototype.
During the whole process, we were driven by the requirements we stated in~\cref{sec:feasibility}. 
In particular, the customization we carried out to manufacture our \SysName prototype is not altering the substantial functioning and design of the earable. 
In fact, in-ear-facing microphones are \textit{already} embedded in commercial, off-the-shelf earbuds (e.g., Apple AirPods Pro)~\cite{airpodspro,honormagicearbuds}.
However, they are mainly used for noise cancellation and, unfortunately, companies do not expose APIs to access in-ear audio recordings.
Therefore, our hardware prototype simply consists of a cheap and easily available in-ear facing microphone placed in a commercially available pair of earbuds. More precisely, we chose a pair of MINISO Marvel earphones~\cite{marvelearphones}. The reasoning behind our choice being: (1) their case is large enough to comfortably fit a Micro-Electro-Mechanical System (MEMS) microphone without the need for removing existing components; (2) the earphones have a removable silicone ear-tip, well suited to occlude people's ear canals. In addition, spare ear-tips of different sizes are also included.
Regarding the microphone, we opted for an analog MEMS SPU1410LR5H-QB~\cite{spu1410}. We selected this particular model given its wide frequency response, spanning flat from 20~Hz to 20~kHz.

\begin{figure}[]
	\centering
	\includegraphics[scale = 0.30]{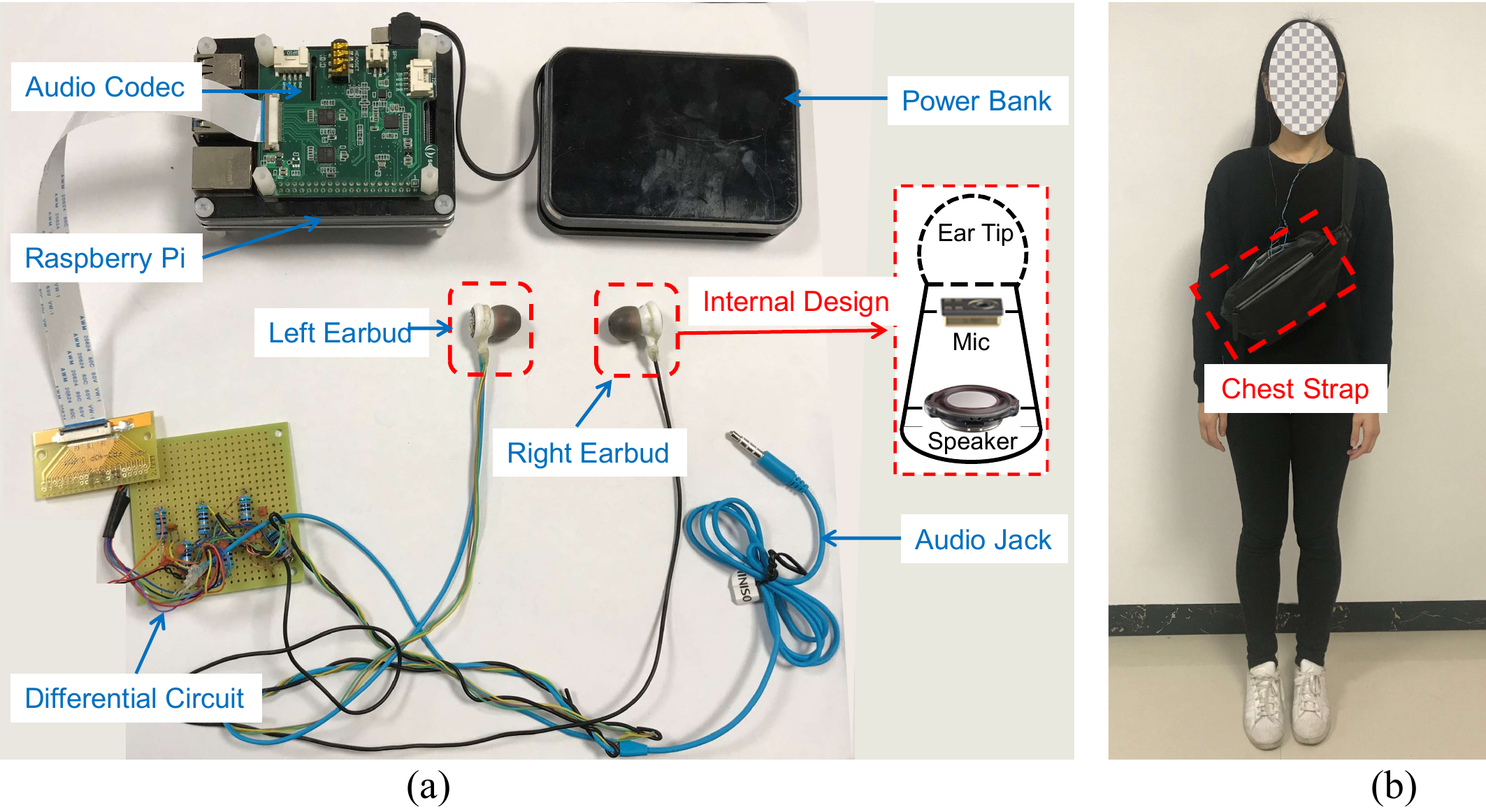}
	\caption{ (a) The designed earbuds prototype and accompanying data recording device, (b) illustration of a participant wearing the device.}
	\vspace{-0.15in}
	\label{fig:device_subject}
\end{figure}

\begin{figure*}[]
	\centering
	\subfigure[OC-SVM]{
	    \centering
	    \includegraphics[scale = 0.26]{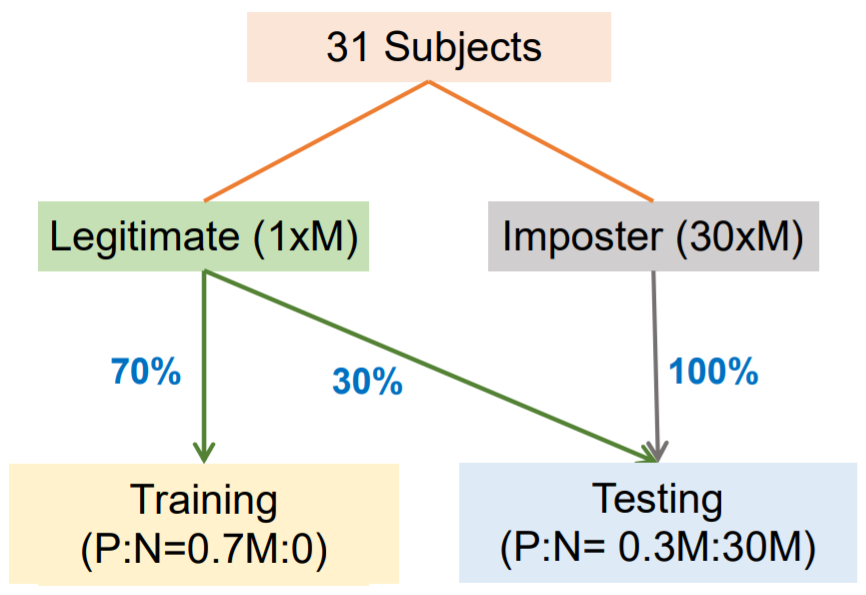}
	    \label{fig:train_test_1}}
    \subfigure[Bi-SVM (Imbalanced)]{
	    \centering
	    \includegraphics[scale = 0.26]{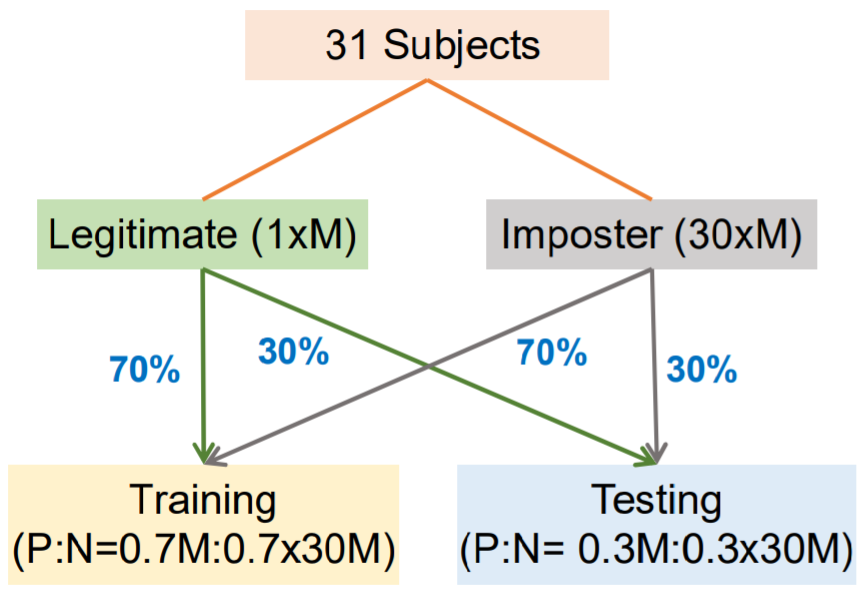}
	    \label{fig:train_test_2}}
	\subfigure[Bi-SVM (Balanced-All)]{
	    \centering
	    \includegraphics[scale = 0.15]{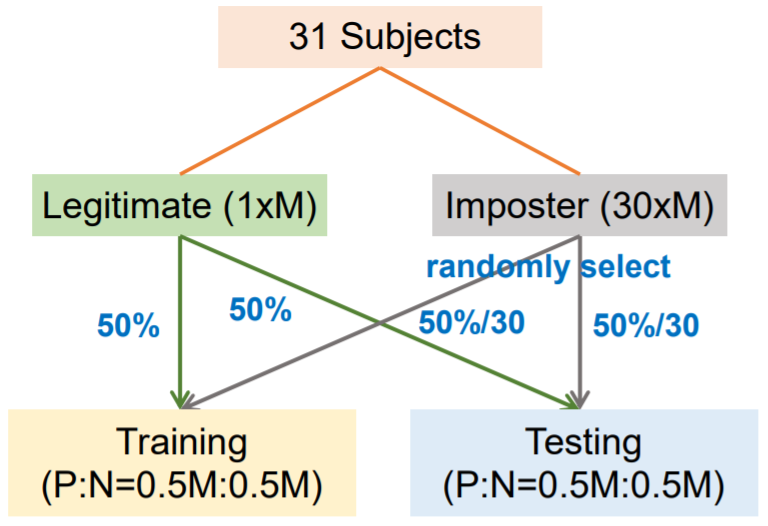}
	    \label{fig:train_test_3}}
	\subfigure[Bi-SVM (Balanced-5)]{
	    \centering
	    \includegraphics[scale = 0.26]{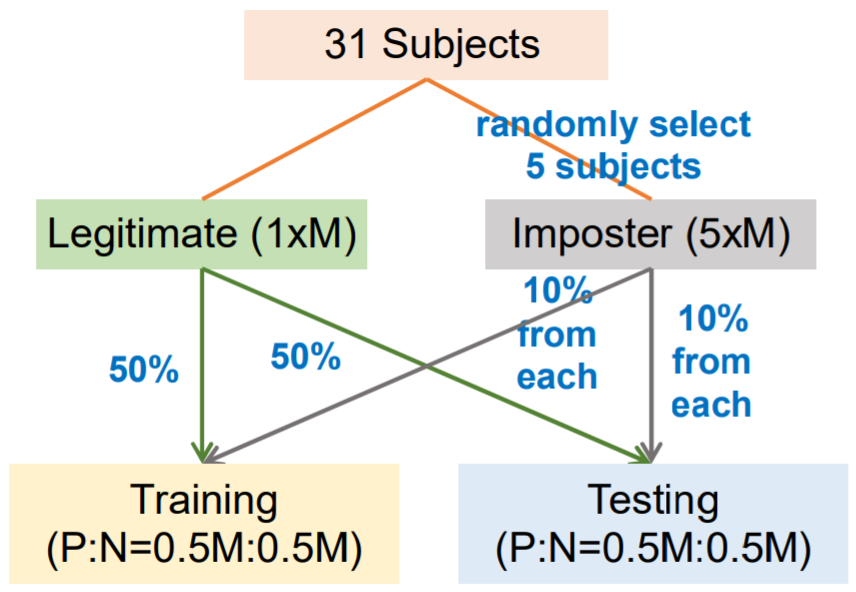}
	    \label{fig:train_test_4}}
    \vspace{-0.1in}
	\caption{Training and testing data splitting scheme for (a) one-class SVM, (b) imbalanced binary SVM, (c) balanced binary SVM with all subjects' data, and (d) balanced SVM with part of subjects' data. $P:N$ represents the ratio between positive (legitimate) and negative (impostor) gait.}
	 \label{fig:training_schemes}
\end{figure*}

\Cref{fig:device_subject} (a) depicts our system in its entirety. In particular, it also reports the internal design of our customized bud, showing how we positioned the microphone immediately in the vicinity of the ear-tip. By doing so, we achieve a higher signal-to-noise ratio (SNR) than what we would have if we had placed the microphone behind the speaker. Notably, this modification does not lead to any deterioration in the audio playback quality. 
To assess that, we recruited 31 subjects and, after reproducing a music piece from both a regular and a modified pair of earbuds, we requested them to provide us with feedback concerning the audio quality. In this phase of the experiment, the subjects were agnostic of presence/absence of the microphone. 
Interestingly, 29 out of 31 people could not experience any difference in audio quality. Surprisingly, 2 even attributed a slightly better quality to the modified bud. We provide more details on our data collection campaign in~\Cref{sec:data-collection}.

In addition to the internal design of our customized earbuds,~\Cref{fig:device_subject} (a) presents the data-logging component we used during our data collection. Both the left and right buds are modified as described above. Notice the speakers in the buds are normally connected to the original 3.5~mm audio jack. 
To achieve the best possible SNR, we connected each microphone to a Differential Circuit right before sampling them with an Audio Codec. 
Specifically, the codec we opted for is an 8-channel audio recording with two ADC chips (AC 108) named ReSpeaker Voice Accessory HAT~\cite{repeakerhat}. The audio codec is controlled by the Python script executed on a Raspberry Pi 4.
The microphone recordings are sampled at 48~kHz. 
Power supply comes from an off-the-shelf power bank, thus allowing us to collect data with the subjects in motion. To reduced the burden of the volunteers when they were moving, we placed all the prototype components (but the earbuds) in a chest strap, as of in~\Cref{fig:device_subject}~(b).

\subsection{Data Collection}\label{sec:data-collection}
After having obtained clearance for carrying out the studies from the Ethics Board of our institution, we recruited 31 subjects. Out of the 31 participants, 15 of them were females and the remainder 16 were males. 
The mean age of the participants was 26.6$\pm$5.8.
All the participants received a voucher in exchange of their time.
After having taken COVID-19 related precautions (i.e.~face masks, hands sanitizing, etc.~), we admitted the participants, one at a time, into the room of the experiment (a 12 $\times$ 6 square meters room).
We instructed the participants to walk in circles following the perimeter walls of the room. The room was quiet, with a noise level of approximately 30~dB. We asked the user to keep their normal pace and walking style. As a gait cycle is composed of two consecutive steps, the subjects always started with their left foot and finished with the right foot when walking. We considered different ground materials, as tiles and carpet, and multiple conditions, with the participants walking barefoot, with slippers, with sneakers, and while speaking. Factoring in all these variables allows us to further assess the robustness, and generalizability, of \SysName.
For each of these conditions, the participants walked continuously for 1.5~minutes (a session). During each walk, the subject counted the number of steps he/she took. This served as ground truth. Walking at normal speed, all the subjects made 156--176 steps per 1.5~minutes-long session. Eventually, each participant performed 8 (2 ground material $\times$ (3 footwear + 1 speaking)) different walking sessions, accounting for a total of 52,046~steps (i.e.~26,023 gait cycles).

\section{Performance Evaluation}\label{sec:eval}
In this section, we present the data collection procedure, the training methodology, as well as different variables we consider while assessing the performance of our system.

\subsection{Metrics}
The metrics we consider to assess the goodness of the proposed \textcolor{black}{identification} system are:
\begin{itemize}
    \item \textbf{False Acceptance Rate} (FAR): a common metric for an \textcolor{black}{identification} system that describes the system's likelihood of \textcolor{black}{successfully identifying} a non-legitimate-user;
    \item \textbf{False Rejecting Rate} (FRR): also known by the name of False Negative Rate (FNR), it indicates the \textcolor{black}{identification} system's likelihood of rejecting the legitimate-user;
    \item \textbf{Balanced Accuracy} (BAC): given we also assess the performance of our system in the case of unbalanced training and testing sets, we consider BAC to gauge the real accuracy of our system. Specifically, BAC is defined as $\frac{TPR+TNR}{2}$, where TPR and TNR are True Positive and True Negative Rate, respectively. The True Positive Rate of an \textcolor{black}{identification} system is its goodness in recognizing legitimate users, whilst True Negative Rate indicates the value of the system in protecting the user from attackers.
\end{itemize}

\subsection{Training-testing protocol}

\begin{figure*}[t]
	\centering
	\subfigure[]{
	    \centering
	    \includegraphics[scale = 0.29]{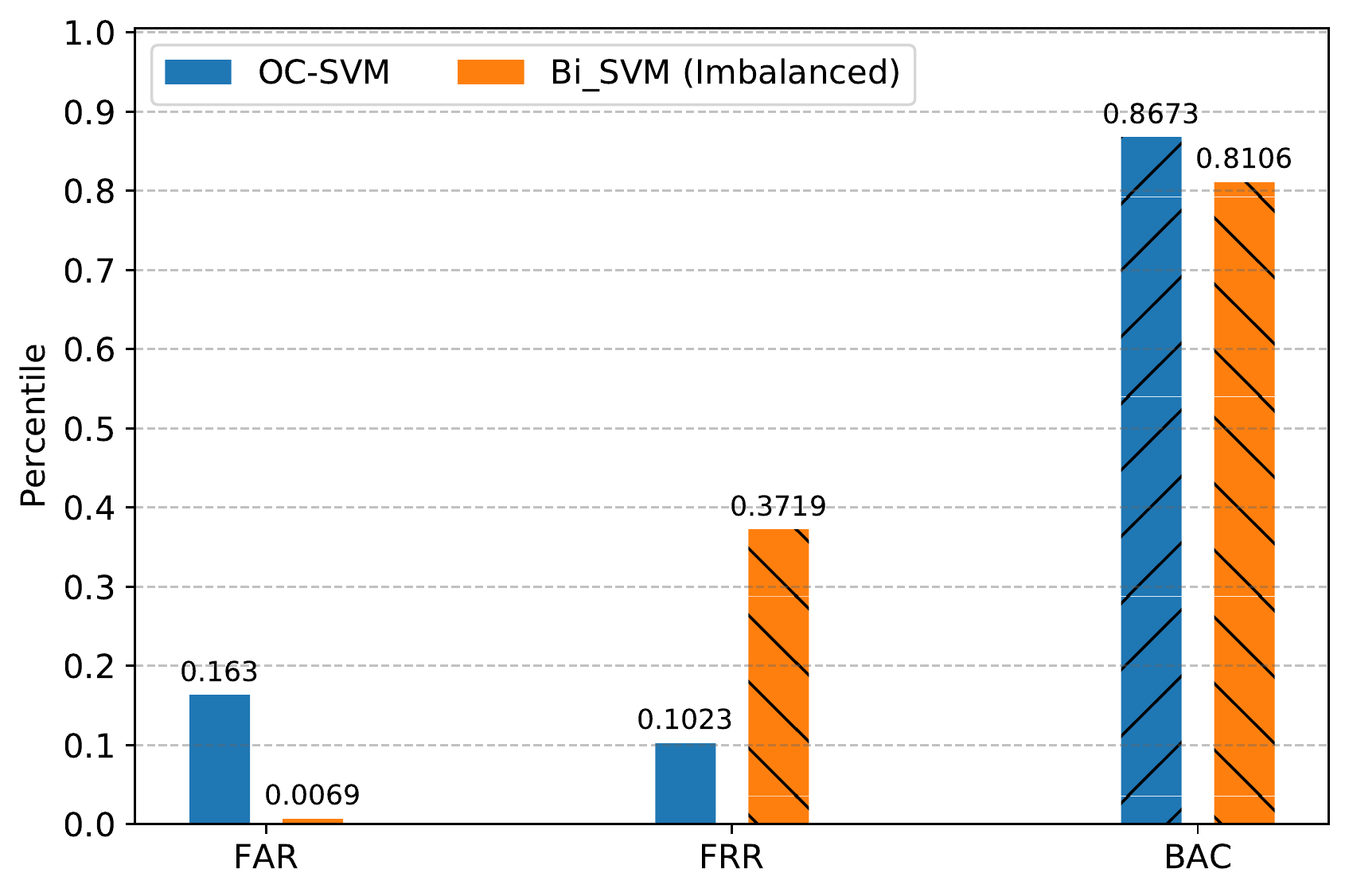}
	    \label{fig:overall_metrics}}
    \subfigure[]{
	    \centering
	    \includegraphics[scale = 0.28]{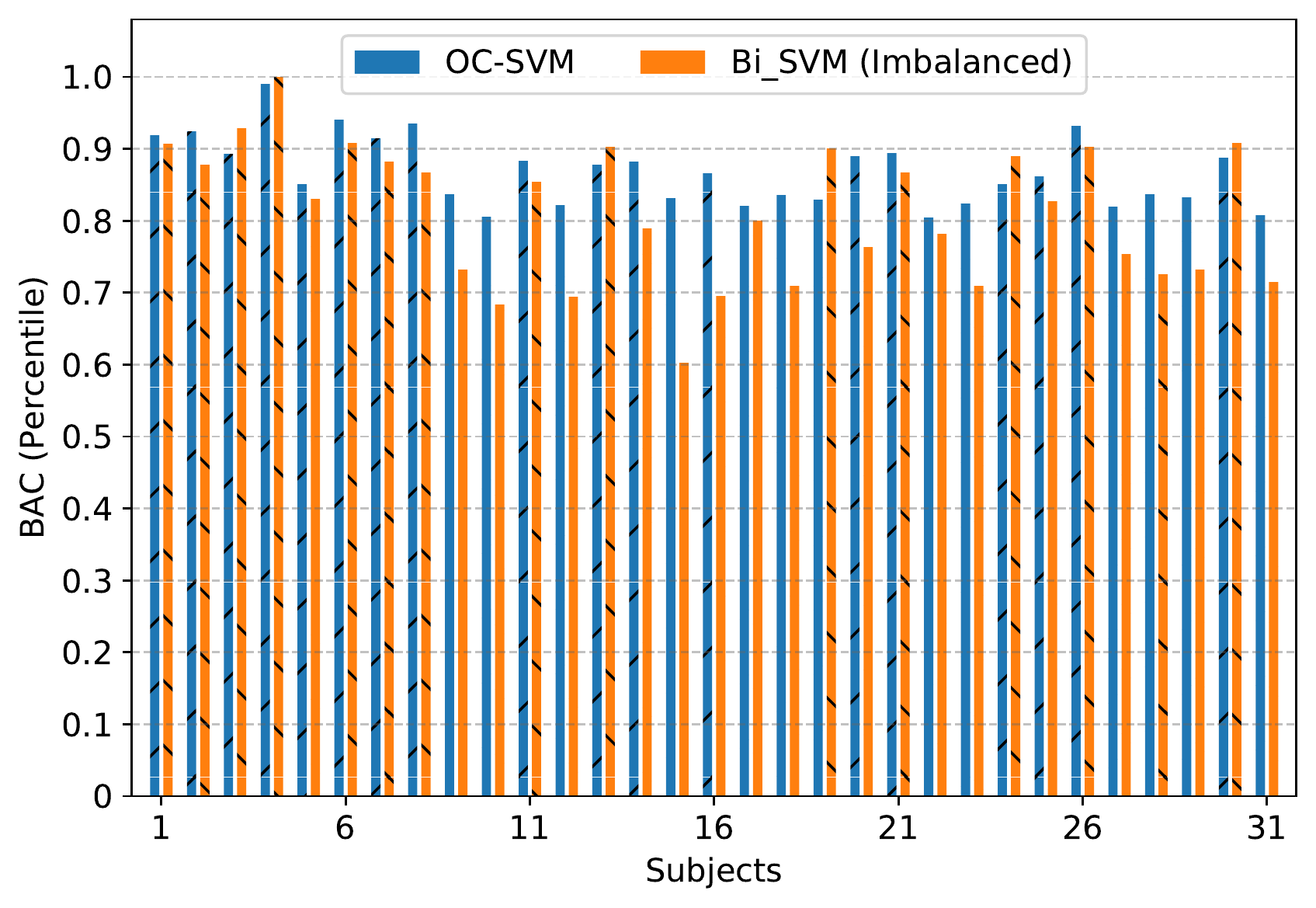}
	    \label{fig:overall_subjects}}
	\subfigure[]{
	    \centering
	    \includegraphics[scale = 0.29]{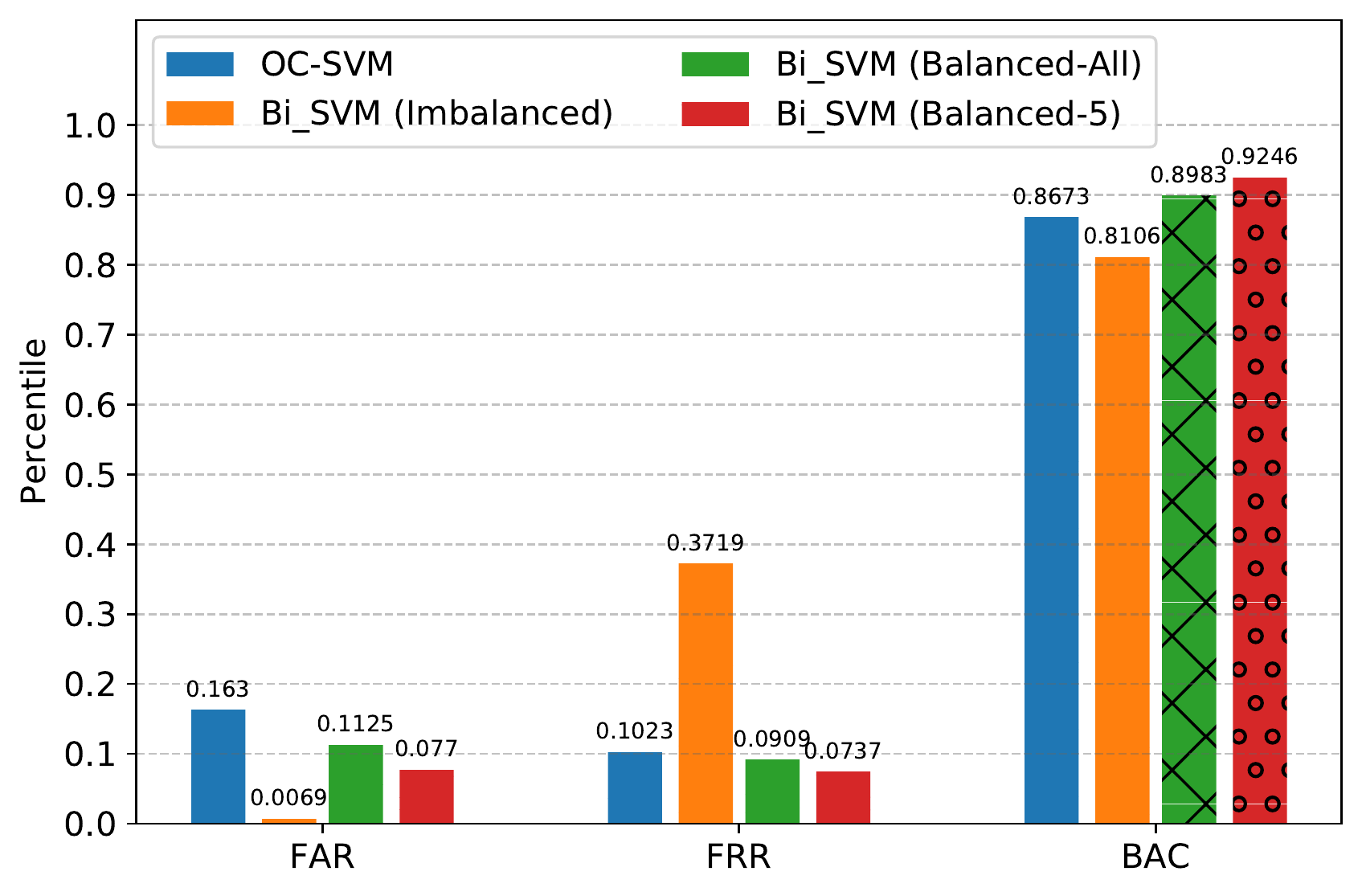}
	    \label{fig:data_imbalance}}
	\vspace{-15pt}
    \caption{(a) Overall performance of \SysName (averaged across 31 subjects), (b) BAC of each individual using OC-SVM and Bi-SVM (Imbalanced). (c) Comparison of the four proposed training-testing protocol, which also reflects the impact of data imbalance.}
	 \label{fig:overall_performance}
\end{figure*}

To evaluate our system, we propose four training-testing protocols, as shown in \Cref{fig:training_schemes}. The number of gait cycles collected from each of the 31 subjects is denoted as $M$. The first two schemes ((a) and (b)) involve different amounts of positive and negative data during training and testing and such data imbalance might affect the performance. Thus, we further propose two protocols ((c) and (d)) that ensure balanced positive and negative samples during training and testing. 
All the four schemes consider adversarial attacks during testing.
Notably, informed by the physiological explanation of gait (\Cref{sec:gait-primer}), we did not review mimic attacks: by being so strongly correlated with the muscular-skeletal structure of the individual, gait is extremely hard to mimic --if not impossible. 
Moreover, the fact that the in-ear (bone-conducted) audio we leverage is further modulated by the skeleton of the subject, constitutes an additional barrier against mimic attacks.

\begin{figure*}[t]
	\centering
	\subfigure[]{
	    \centering
	    \includegraphics[scale = 0.29]{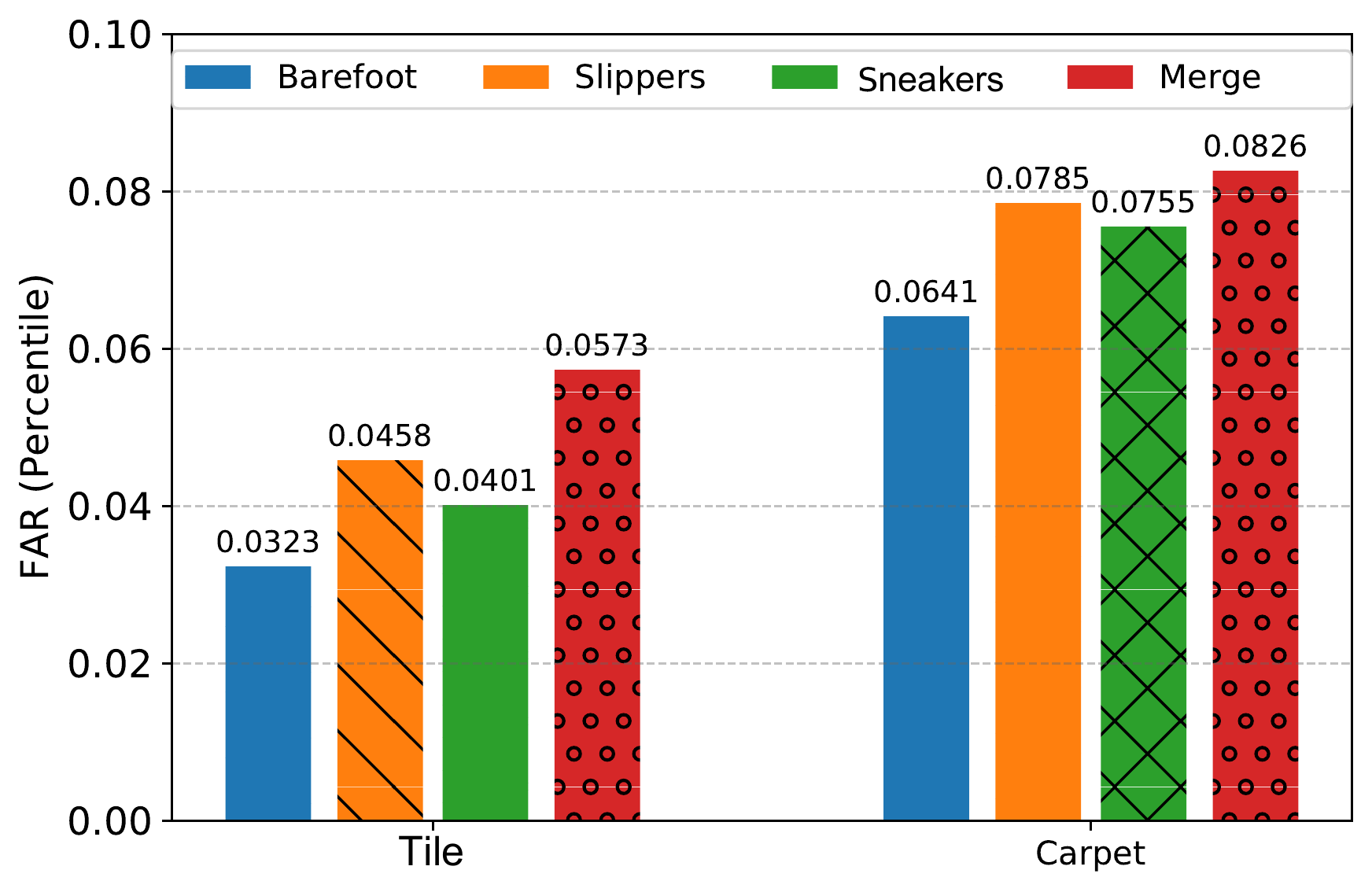}
	    \label{fig:shoe_ground_far}}
	\subfigure[]{
	    \centering
	    \includegraphics[scale = 0.29]{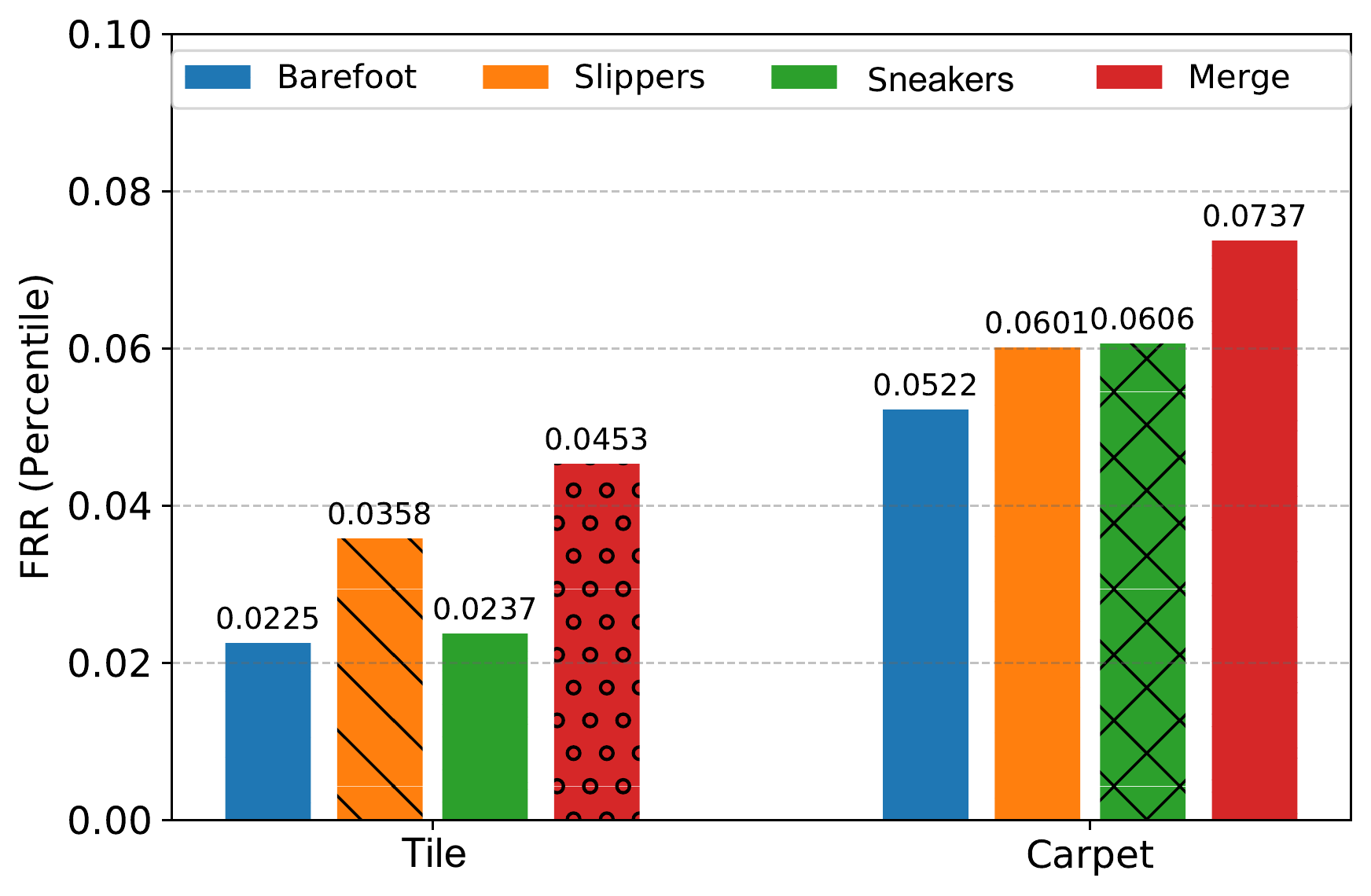}
	    \label{fig:shoe_ground_frr}}
    \subfigure[]{
	    \centering
	    \includegraphics[scale = 0.29]{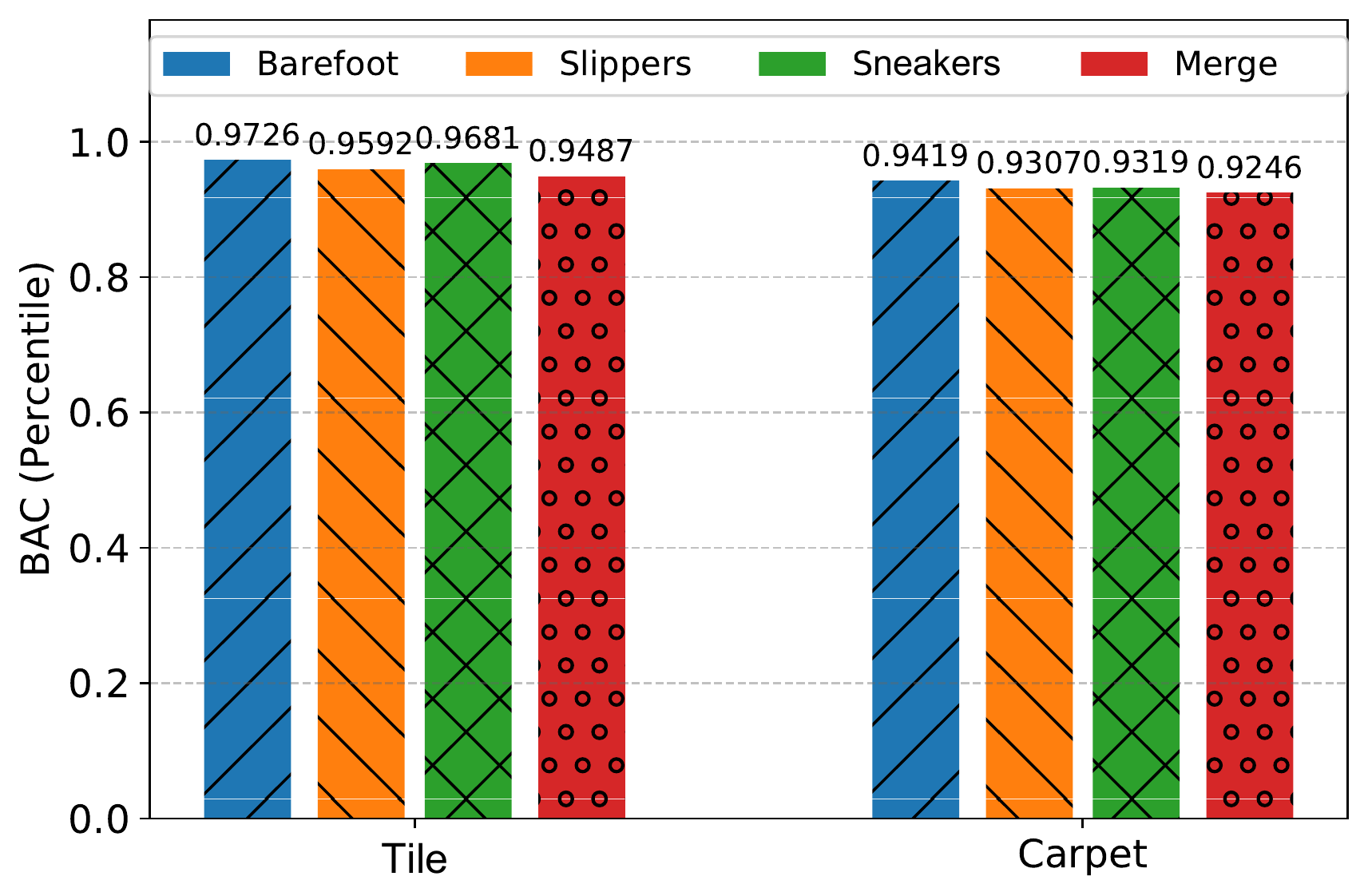}
	    \label{fig:shoe_ground_bac}}
	\vspace{-15pt}
	\caption{Performance comparison (a) FAR, (b) FRR, and (c) BAC, of different footwear and ground material.}
	 \label{fig:shoe_ground_impact}
\end{figure*}

\begin{itemize}
    \item \textbf{(a)} One-Class SVM (\textbf{OC-SVM}): one subject is iteratively selected as the legitimate user and the rest are regarded as impostors.
    Training dataset only consists of 70\% ($0.7 \times M$) data from the legitimate user, and the testing dataset is composed of 30\% ($0.3 \times M$) legitimate user data and all impostor data ($30 \times M$). 
    \item \textbf{(b)} Imbalanced Binary SVM (\textbf{Bi-SVM (Imbalanced)}): one subject is iteratively selected as the legitimate user and the rest are regarded as impostors. Training dataset consists of 70\% legitimate
    user's data ($0.7 \times M$) and 70\% impostors' data ($30 \times 0.7 \times M$), and the testing dataset is composed of 30\% ($0.3 \times M$) legitimate user's data and 30\% ($30 \times 0.3 \times M$) impostors' data.
    \item \textbf{(c)} Balanced Binary SVM with all subjects' data (\textbf{Bi-SVM (Balanced-All)}): one subject is iteratively selected as the legitimate user and the rest are regarded as impostors. Training dataset consists of 50\% ($0.5 \times M$) legitimate user's data and the same number ($0.5 \times M$) of gait cycles that are randomly selected from the 30 impostors. The testing dataset is composed of the rest 50\% ($0.5 \times M$) legitimate user's data and another $0.5\times M$ randomly selected impostor data.
    \item \textbf{(d)} Balanced Binary SVM with part of subjects' data (\textbf{Bi-SVM (Balanced-5)}): one subject is iteratively selected as the legitimate user and five (5/30) subjects are randomly selected as impostors. Training dataset consists of 50\% ($0.5 \times M$) legitimate user's data and 10\% from each of the five impostors ($5 \times 0.1 \times M$). The testing dataset is composed of the rest 50\% ($0.5 \times M$) legitimate user's data and another 10\% ($5 \times 0.1 \times M$) data from each of the impostors.
\end{itemize}

In the remainder of this section, we present our experimental results under various conditions.

\subsection{Overall Performance}

We first evaluate the overall performance of \SysName by combining all the collected data together, i.e., different ground material and footwear.  \Cref{fig:overall_metrics} presents the results obtained with the first two training protocols (OC-SVM and Bi-SVM (Imbalanced)), which are averaged over the 31 subjects. We can observe that with both methods, \SysName can achieve over 80\% balanced accuracy (BAC). The FAR and FRR are relatively high, which might be caused by the imbalanced dataset and the variability of different waking conditions. We will discuss this impact in the following sub-sections. \Cref{fig:overall_subjects} plots the BAC of each individual with the two protocols. We can see that although the performance varies among subjects, most of the subjects achieve over 80\% BAC. In addition, we found that the best training-testing protocol is subject-dependent as some users obtain higher BAC with OC-SVM, while others achieve better performance with Bi-SVM (Imbalanced). Thus, it is necessary to optimize the training protocol for each individual.

\subsection{Impact of Data Imbalance}
Intuitively, Bi-SVM is expected to perform better than OC-SVM. The reason is that OC-SVM is similar to a clustering problem and the model has to learn the correlation of the data without information on outliers.
While for Bi-SVM, the inclusion of benchmark (impostor) data provides additional information about the negative samples, so that the model can learn a more accurate and converged representation of the legitimate user. 
However, Bi-SVM performs worse~\Cref{fig:overall_metrics}.
Specifically, FRR is dramatically higher than FAR. Such phenomenon is also observed in~\cite{zou2018bilock} and we suspect this originates from the imbalanced positive and negative samples during training and testing. Thus, we re-run the experiments using the proposed two training schemes with balanced samples.

As shown in~\Cref{fig:data_imbalance}, the large difference between FAR and FRR disappears when the number of positive and negative classes is balanced. As expected, Bi-SVM (Balanced-All) and Bi-SVM (Balanced-5) obtains better performance (lower FAR/FRR and higher BAC), with BAC significantly enhanced, from 81\% to 92.5\%. Bi-SVM (Balanced-5) even performs slightly better as the impostor data is also balanced (10\% from each). We therefore only report the results obtained with Bi-SVM (Balanced-5) in the rest of the evaluation.

\begin{figure*}[t]
	\centering
	\subfigure[]{
	    \centering
	    \includegraphics[scale = 0.29]{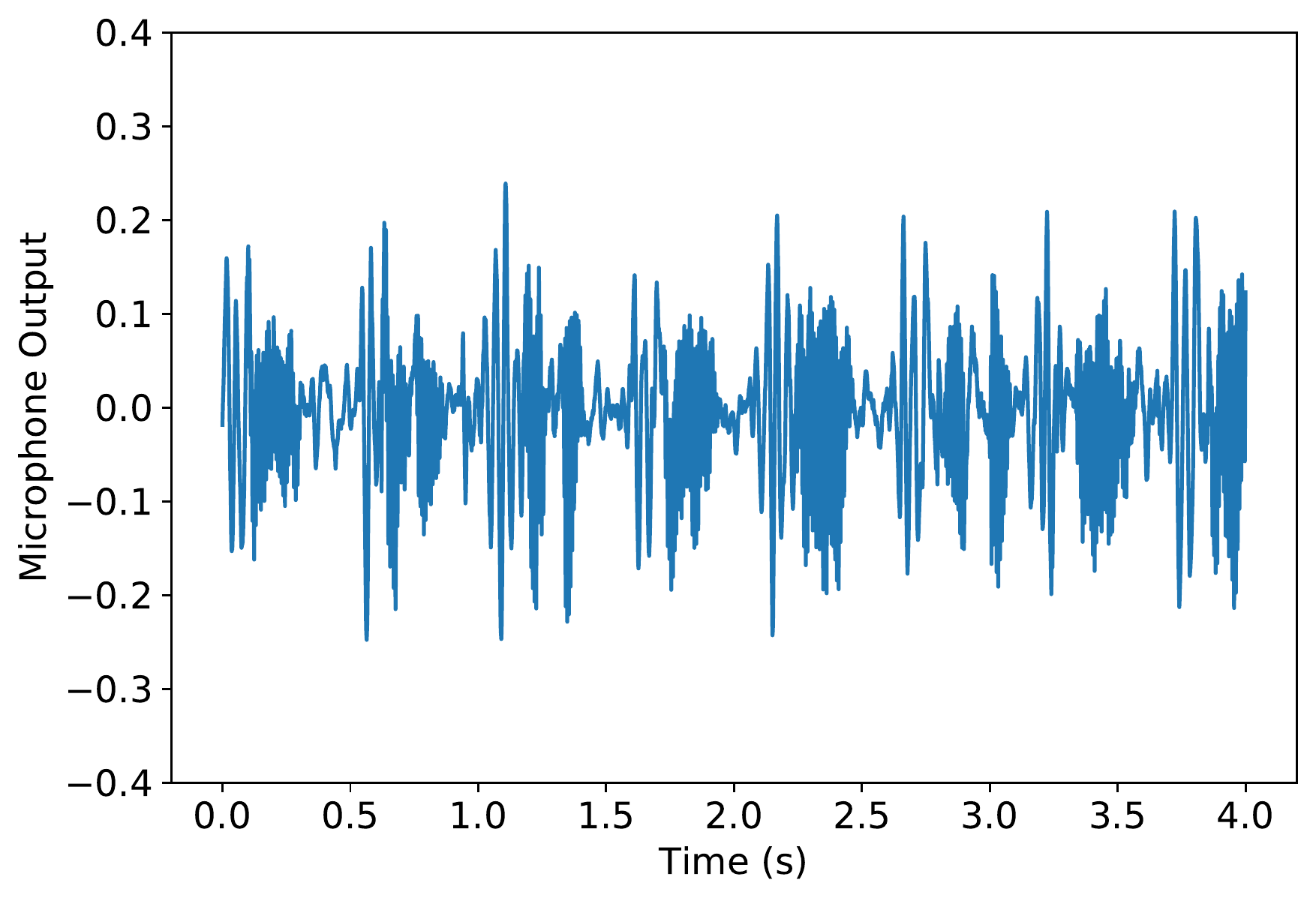}
	    \label{fig:brick_speak_original}}
	\subfigure[]{
	    \centering
	    \includegraphics[scale = 0.29]{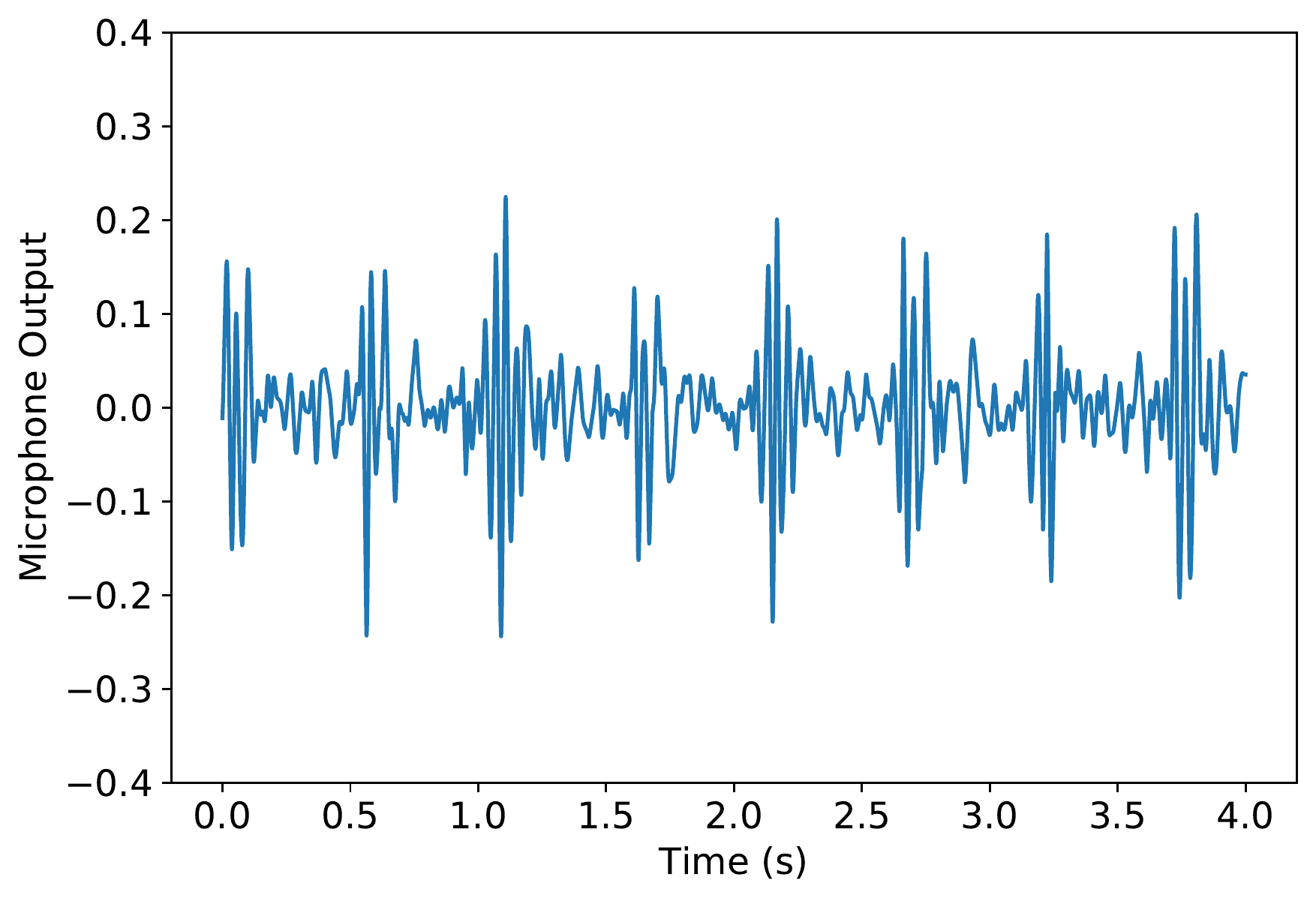}
	    \label{fig:brick_speak_filtered}}
    \subfigure[]{
	    \centering
	    \includegraphics[scale = 0.29]{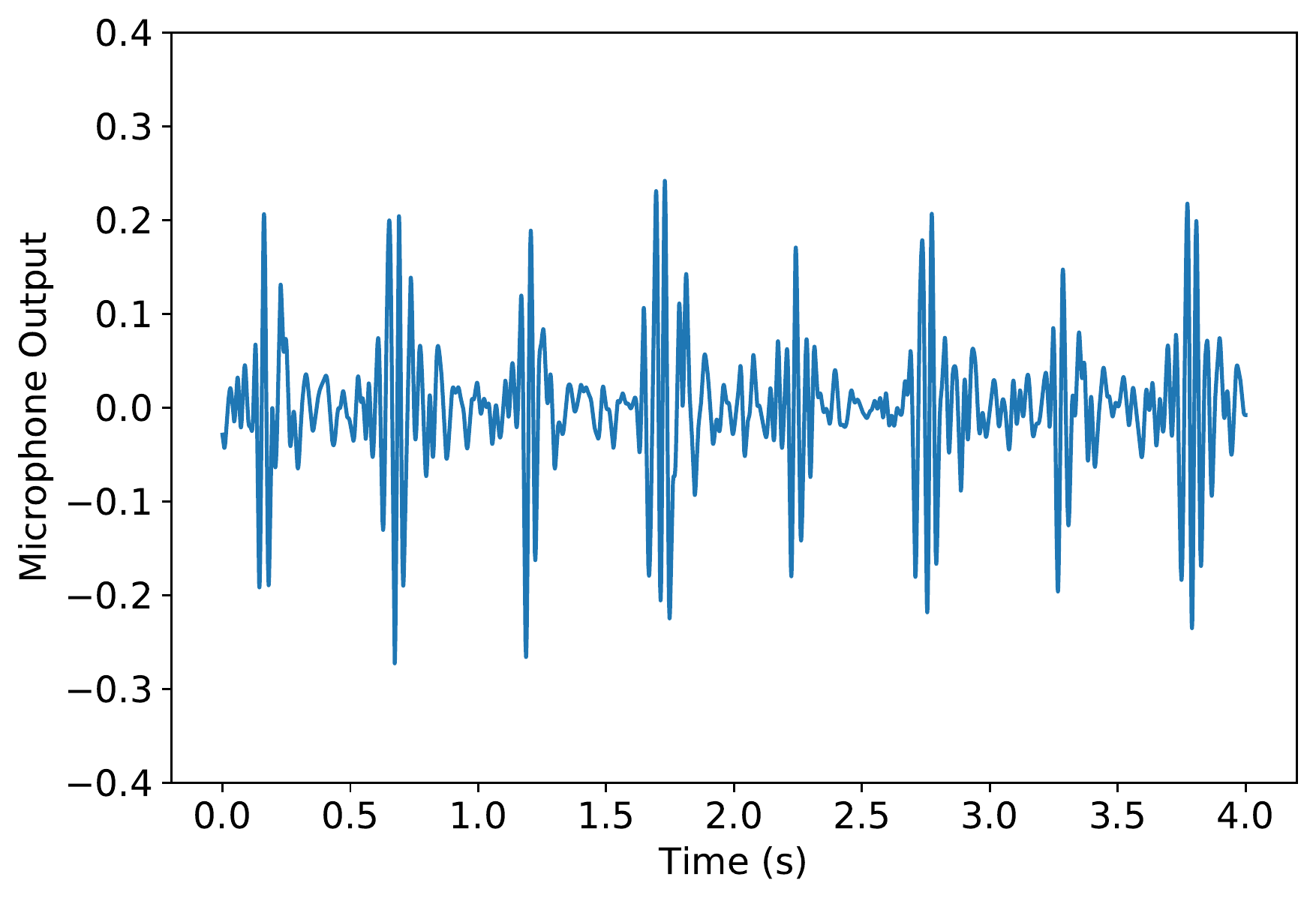}
	    \label{fig:brick_no_speak}}
	\vspace{-15pt}
	\caption{Impact of user speaking. We take one subject walking on tiles as an example, (a) original signal with speaking, (b) low-pass filtered signal with speaking, (c) filtered signal without speaking.}
	\vspace{-10pt}
	 \label{fig:impact_speaking}
\end{figure*}

\begin{figure*}[t]
	\centering
	\subfigure[]{
	    \centering
	    \includegraphics[scale = 0.29]{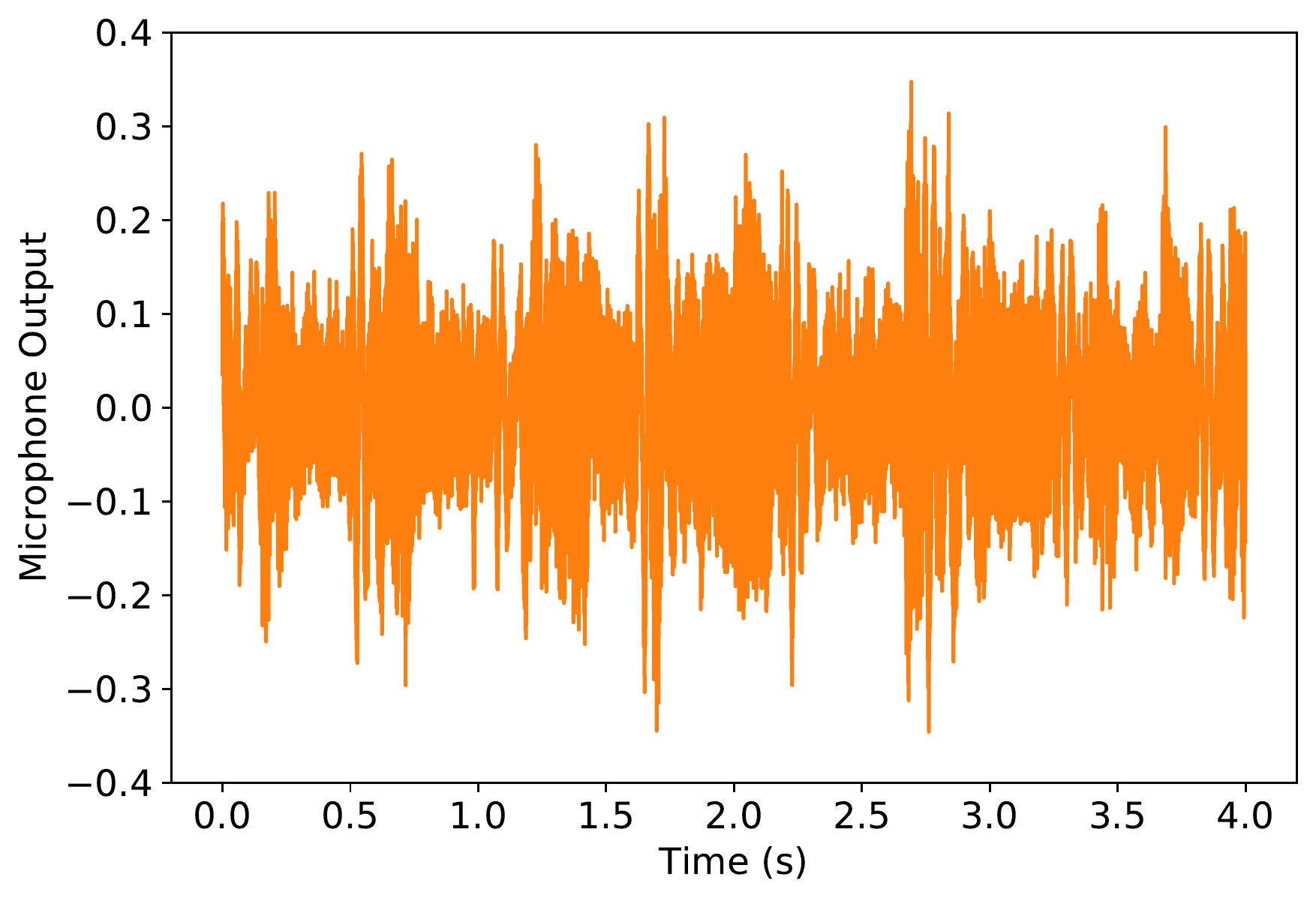}
	    \label{fig:music_original}}
	\subfigure[]{
	    \centering
	    \includegraphics[scale = 0.29]{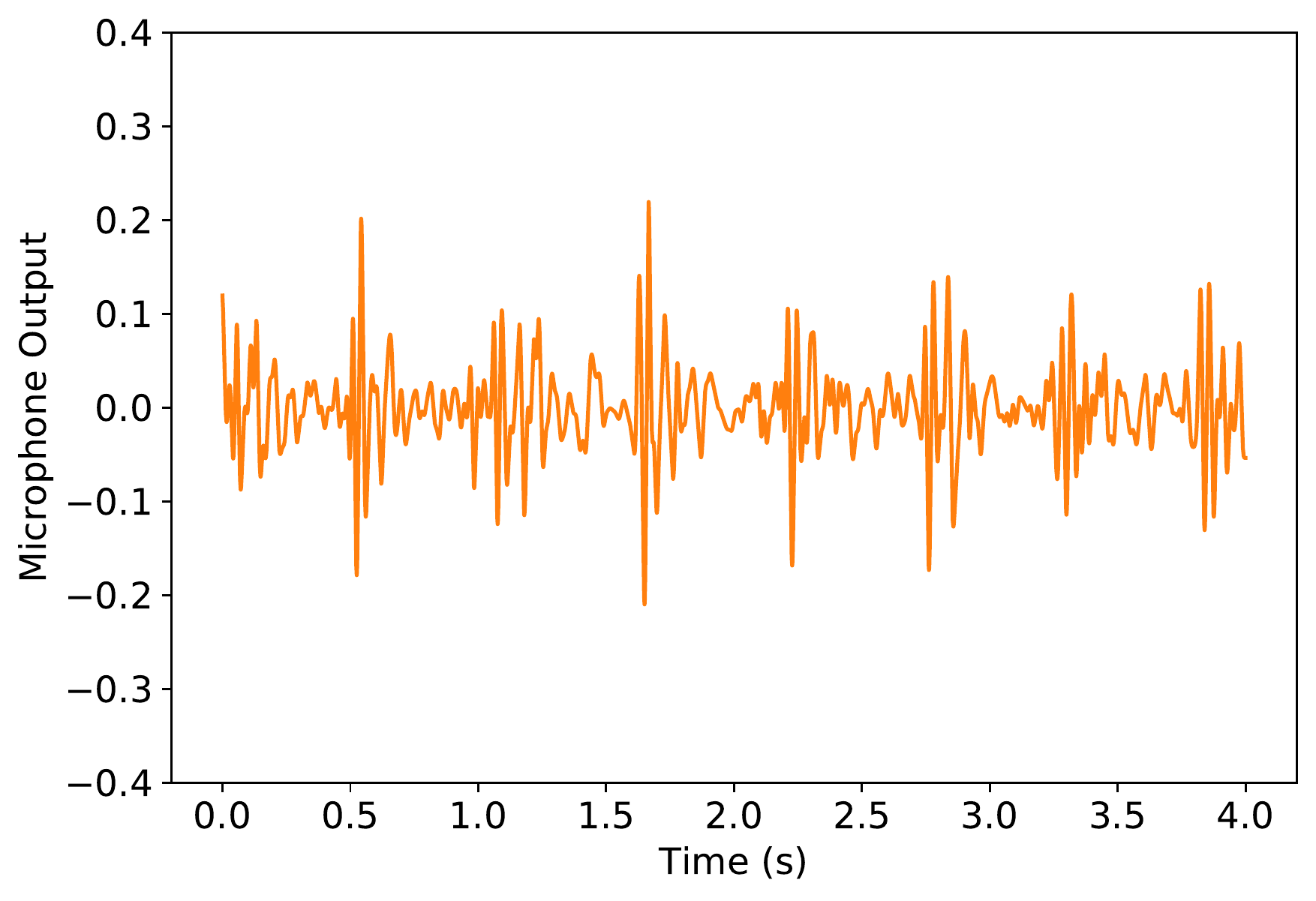}
	    \label{fig:music_filtered}}
    \subfigure[]{
	    \centering
	    \includegraphics[scale = 0.29]{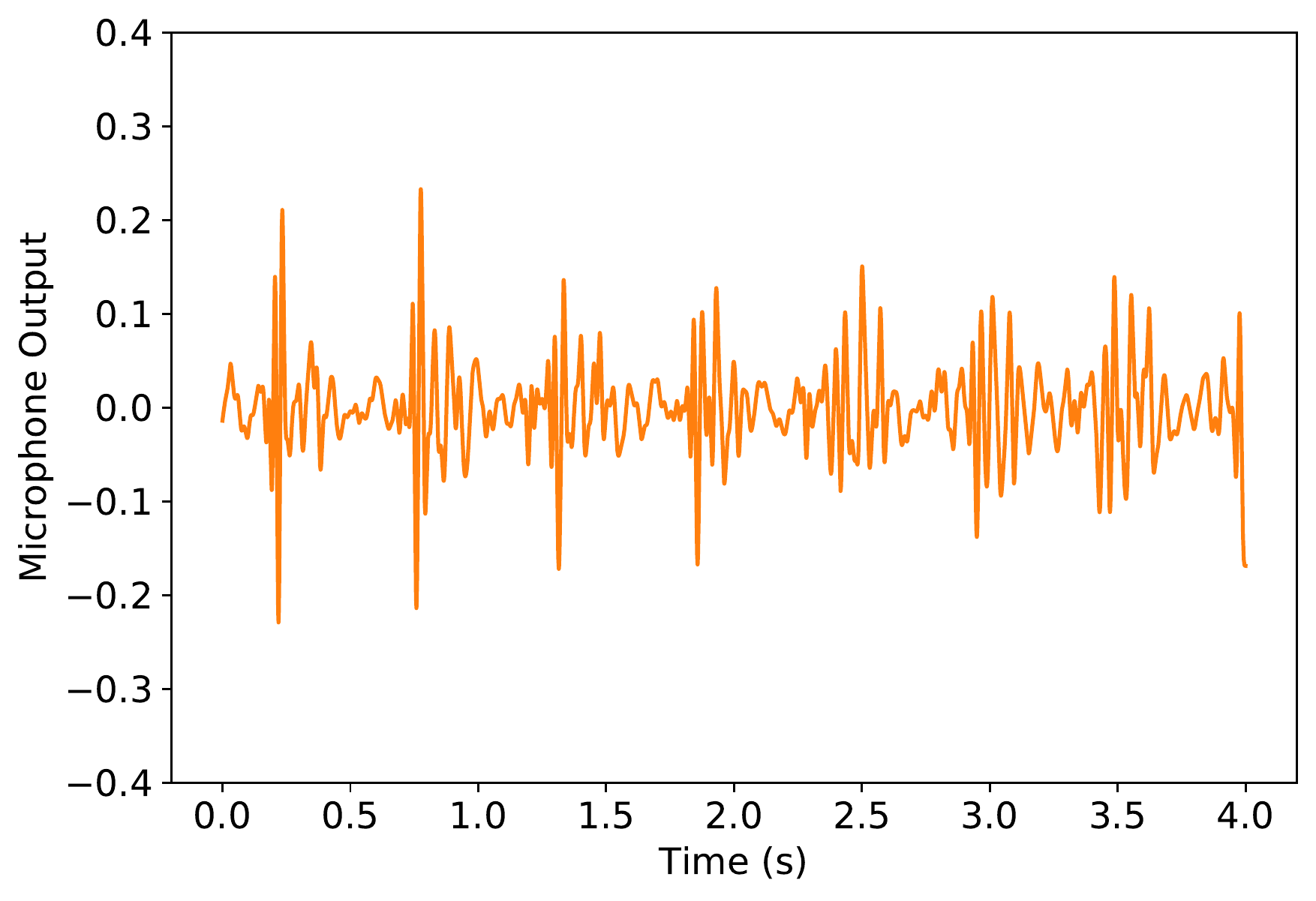}
	    \label{fig:brick_no_music}}
	\vspace{-15pt}
	\caption{Impact of music playback. We take one subject walking on tiles as an example, (a) original signal with music playing, (b) low-pass filtered signal with music playing, (c) filtered signal without music.}
	\vspace{-5pt}
	 \label{fig:impact_music}
\end{figure*}

\subsection{Impact of Walking Condition}
The walking conditions, such as different footwear or ground material, might affect the gait \textcolor{black}{identification} performance. For instance, compared to tiles ground, walking on carpet will result in longer stance phase and weaker vibrations when hitting the ground. These conditions also introduce variations on human gait and therefore making the \textcolor{black}{identification} task tougher. Next, we investigate the impact of footwear and ground material.

\subsubsection{Footwear}
\Cref{fig:shoe_ground_impact} shows the FAR, FRR, and BAC with data collected when the subjects were barefoot, wearing slippers and sneakers, as well as the combination of all these data, respectively. The results indicate that \SysName works well regardless of footwear, with BAC higher than 94\%. Particularly, walking barefoot achieves the best performance, followed by wearing sneakers, whilst wearing slippers is the most challenging case, as slippers introduce more variations during walking. This observation is applicable to both the dataset collected on tiles and carpet. In addition, when evaluating the data collected in a single session, the BAC significantly increased from 92.46\% (\Cref{fig:data_imbalance}) to 97.26\%. 

\subsubsection{Ground Material}
Next, we explore the impact of ground material on the \textcolor{black}{identification} performance. As shown in~\Cref{fig:shoe_ground_impact}, both tiles
and carpet achieve good performance, with BAC higher than 93\%. In particular, for the same footwear (e.g.~sneakers), tiles always obtains better performance (3\% improvement on BAC) compared to carpet. This might because soft carpet counteracts part of the generated vibrations, and therefore the signal-to-noise ratio (SNR) is lower.    

\subsection{Impact of Human Speech}
\label{sec:impact_human_speech}
Human speech might have an impact on the proposed \textcolor{black}{identification} system. On one hand, the voice produced when people speak will be captured by the in-ear microphone, thereby polluting the recorded acoustic gait data. On the other hand, during speaking, the movement of mouth and jaw modifies the structure of human body. Consequently, the generated vibrations will experience a slightly different propagation path, resulting in different modulations on the gait signal. Thus, we asked the subjects to speak when walking on tiles with sneakers (the most common case for daily walking) and explore whether the performance would be affected.

\Cref{fig:impact_speaking} plots the raw microphone data from the left earbud. It is clear that the gait signal is polluted by high-frequency human speech. Fortunately, the actual gait signal we are interested in can be distinguished in frequency domain as \textbf{(1)} the generated vibrations are in low frequencies and \textbf{(2)} the occlusion effect mainly emphasizes the low-frequency components of the bone-conducted sound. \Cref{fig:brick_speak_filtered} illustrates the low-pass filtered version of \Cref{fig:brick_speak_original}. In addition, we plot the filtered signal of the same participant walking without speaking in~\Cref{fig:brick_no_speak}. Visually, the two filtered signals look quite similar and the impact of human speech has been completely removed. Using the Bi-SVM (Balanced-5) protocol, we run the experiments on the collected dataset with users speaking during walking. The results are satisfactory for both tiles (FAR=7.73\%, FRR=4.37\%, BAC=93.95\%) and carpet (FAR=10.73\%, FRR=6.98\%, BAC=91.15\%), denoting how human speech has little impact on \SysName.

\begin{figure*}[]
	\centering
	\subfigure[]{
	\includegraphics[width = 0.29\linewidth]{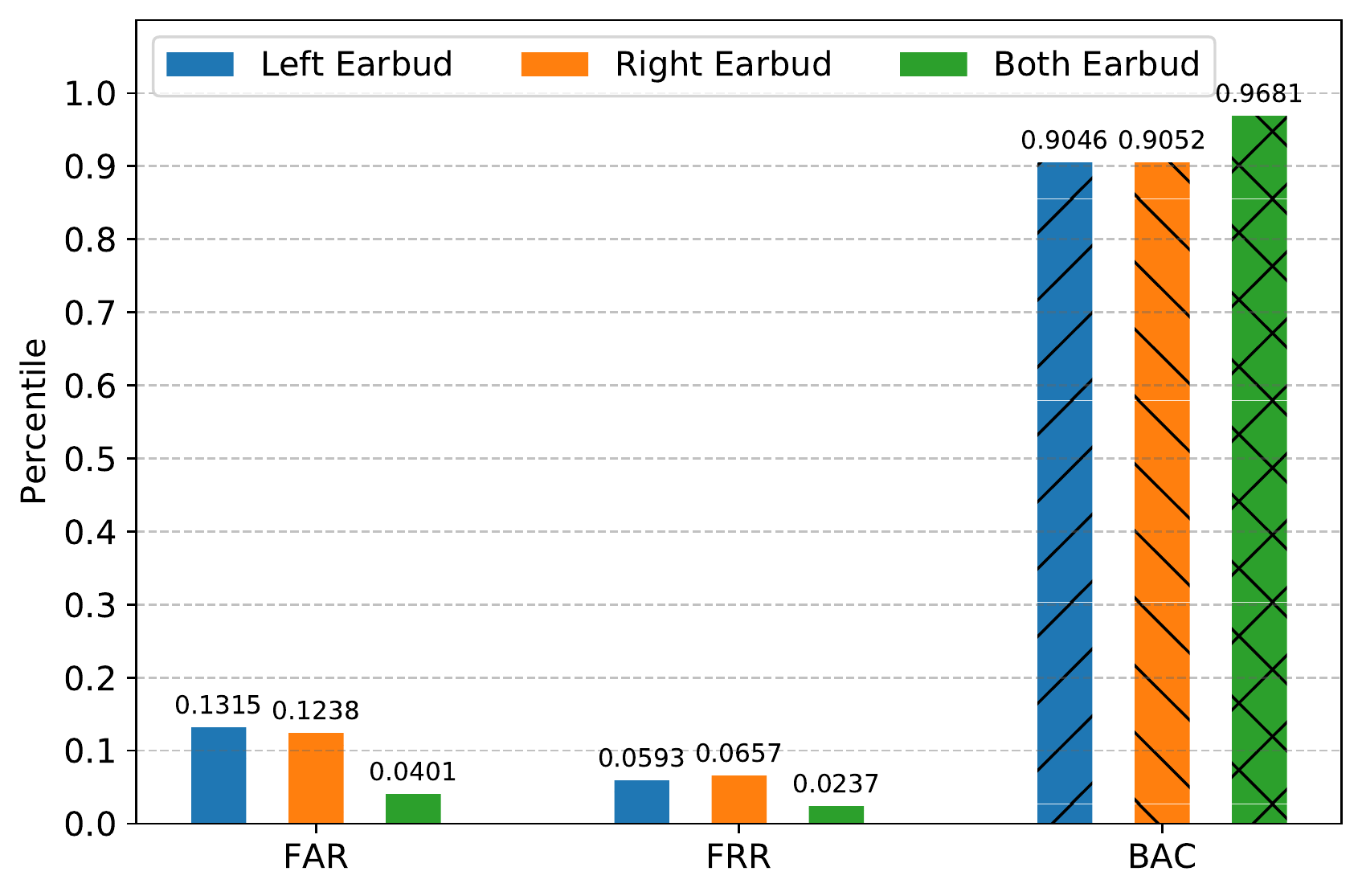}
	\label{fig:left_right_impact}}
\centering
\subfigure[]{
	\centering
	\includegraphics[width = 0.295\linewidth]{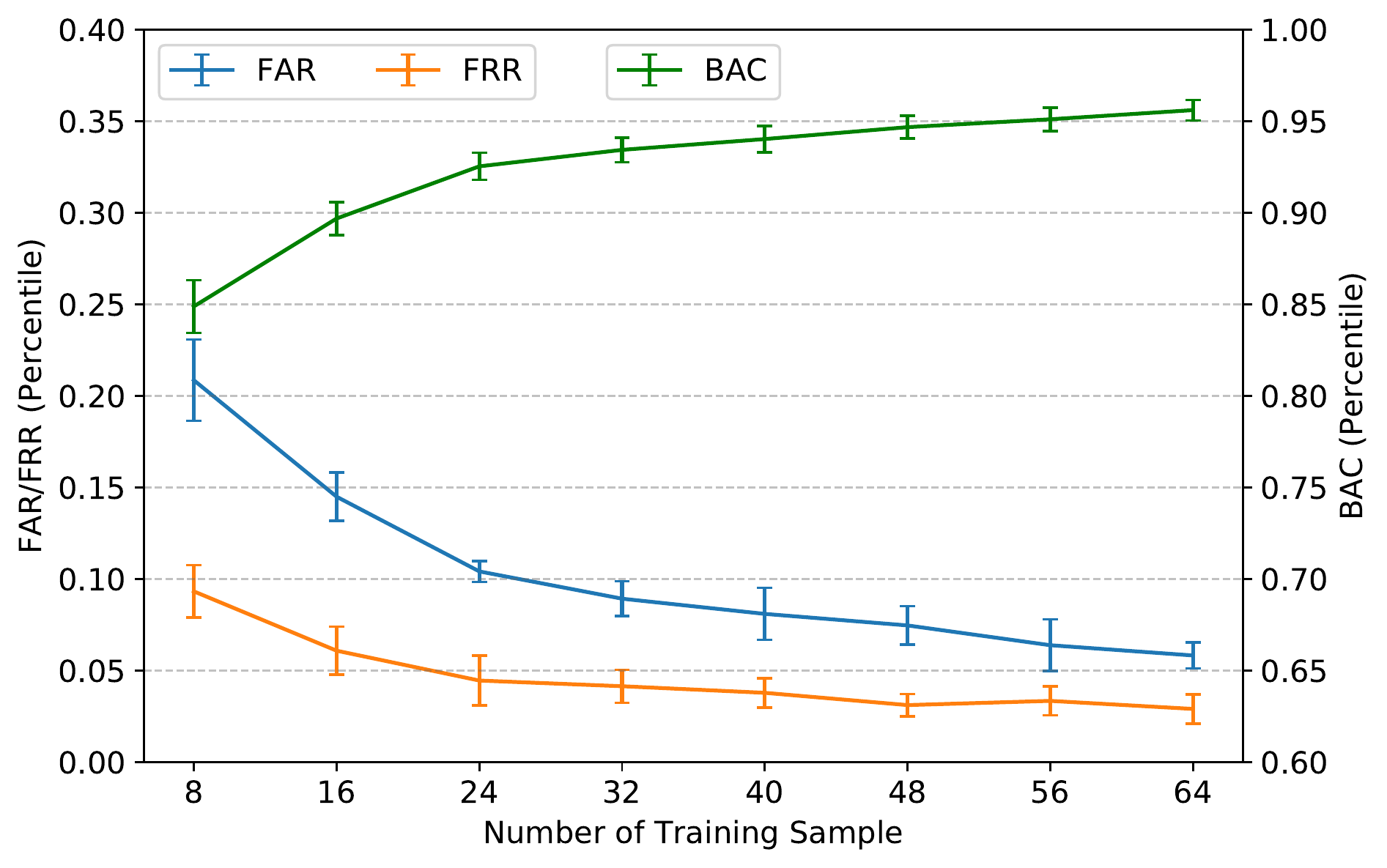}
	\label{fig:training_size}}
\subfigure[]{
	\centering
	\includegraphics[width = 0.29\linewidth]{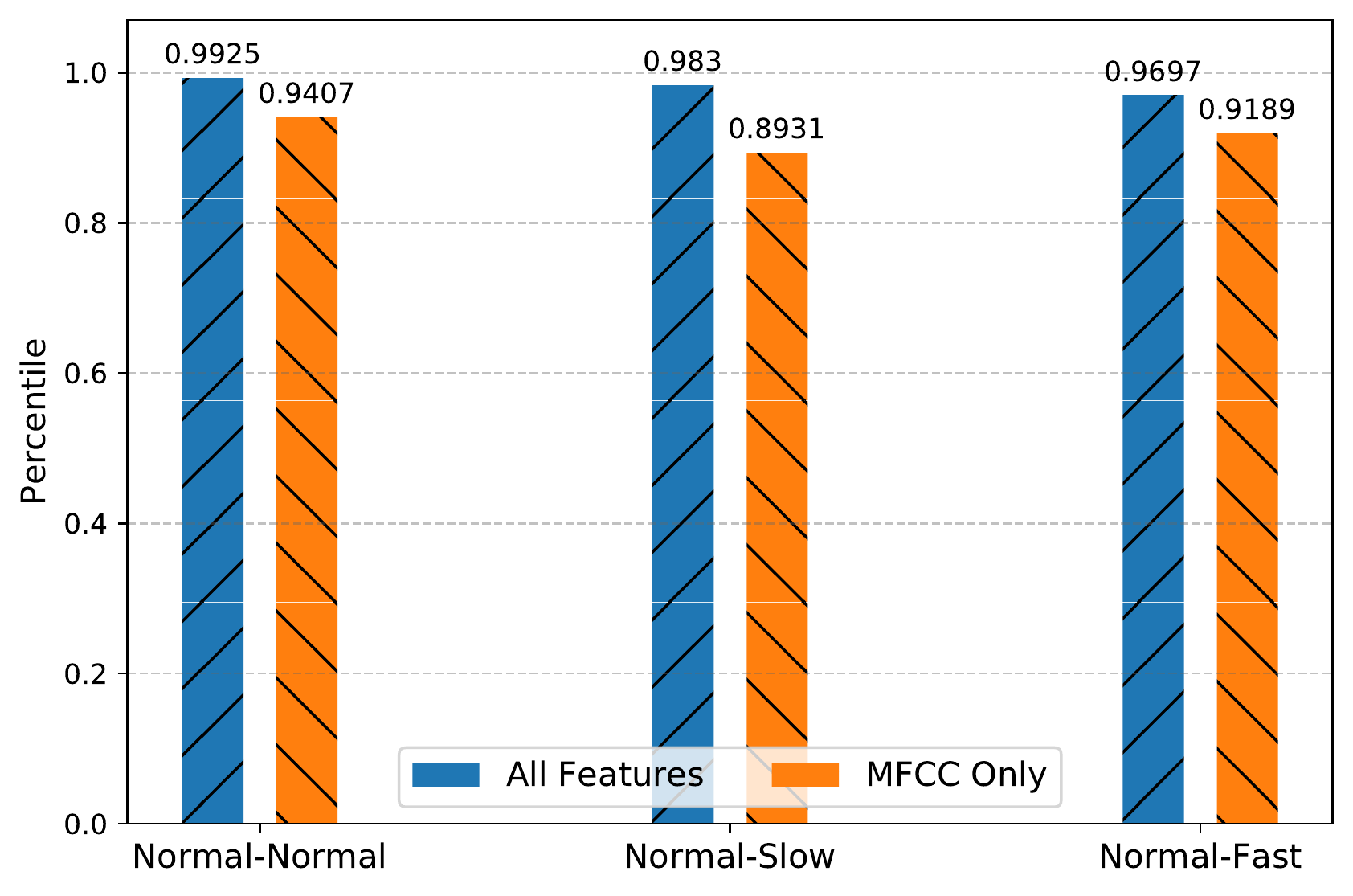}
	\label{fig:speed_feature}}
	\vspace{-15pt}
	\caption{Impact of (a) fusing data from both the earbuds, (b) different training size, and (c) different pace and features, on FAR, FRR, and BAC.}
	\vspace{-5pt}
\end{figure*}

\subsection{Impact of Music Playback}
\label{sec:impact_music_playback}
The general purpose of earbuds is leisure/entertainment, particularly music playback. Due to the vicinity (less than 1 centimeter) of the speaker and in-ear microphone, we investigated whether music playback introduces interference in the gait \textcolor{black}{identification} system. To explore that, we asked one subject to walk when the earbud was playing music at an appropriate volume. \Cref{fig:music_original} plots the original signal from the left earbud, where the target gait signal is overwhelmed by the music. Similar to human speech, these audible sounds are in higher frequencies, while \SysName captures gait in low frequencies (<50~Hz). Thus, after applying the low-pass filter, music can be eliminated and gait signal of interest is clearly observed in~\Cref{fig:music_filtered}. Also, we plot a trace (low-pass filtered) of the same subject walking without playing the music in~\Cref{fig:brick_no_music}, which shows high similarity with signals in~\Cref{fig:music_filtered}.

To further demonstrate how \SysName is robust against different types of music, we conducted a spectrum analysis of the All-time Top 100 Songs launched by Billboard~\cite{top100songs}. Specifically, we applied fast Fourier transform (FFT) analysis on each song and obtain the signal energy at each frequency band. Then, the energy of frequency band lower than 50~Hz was summed and we calculated the energy ratio by $E_{<50}/E_{total}$ ($E_{<50}$ refers to the total energy at frequencies below 50~Hz and $E_{total}$ refers to the total energy of the song). Averaged among the 100 songs, only 1.5\% energy is distributed under 50~Hz, suggesting that the impact of music playback is negligible for \SysName. We also considered the impact of phone calls. Evidence from the literature shows that the frequency range of human voice over telephony transmission is within 300-3400~Hz~\cite{esteban19789}, which can be easily removed after low-pass filtering. Thus, \SysName is compatible with the general purpose of earbuds, without introducing any mutual interference.
Besides, given many existing earbuds already features in-ear microphones for active noise cancellation, EarGate can be operated concurrently. Specifically, the data from in-ear microphones can be delivered to two independent pipelines: noise cancellation algorithm to actively generate the anti-noise and user identification framework to recognize the user.

\begin{table*}[]
\small
\centering
\caption{\textcolor{black}{Identification performance improvements achieved with transfer learning.}}
\begin{tabular}{|
>{\columncolor[HTML]{FFFFFF}}c |
>{\columncolor[HTML]{FFFFFF}}c |
>{\columncolor[HTML]{FFFFFF}}c |
>{\columncolor[HTML]{FFFFFF}}c |
>{\columncolor[HTML]{FFFFFF}}c |
>{\columncolor[HTML]{FFFFFF}}c |
>{\columncolor[HTML]{FFFFFF}}c |
>{\columncolor[HTML]{FFFFFF}}c |}
\hline
\cellcolor[HTML]{FFFFFF} &
  \cellcolor[HTML]{FFFFFF} &
  \multicolumn{3}{c|}{\cellcolor[HTML]{FFFFFF}\textbf{SVM}} &
  \multicolumn{3}{c|}{\cellcolor[HTML]{FFFFFF}\textbf{XGBoost}} \\ \cline{3-8} 
\multirow{-2}{*}{\cellcolor[HTML]{FFFFFF}\textbf{Tasks}} &
  \multirow{-2}{*}{\cellcolor[HTML]{FFFFFF}\textbf{Type of Features}} &
  \textbf{FAR} &
  \textbf{FRR} &
  \textbf{BAC} &
  \textbf{FAR} &
  \textbf{FRR} &
  \textbf{BAC} \\ \hline
\cellcolor[HTML]{FFFFFF} &
  \textbf{VGGish embeddings} &
  0,730 &
  0,231 &
  0,520 &
  0,093 &
  0,092 &
  \textbf{0,907} \\ \cline{2-8} 
\cellcolor[HTML]{FFFFFF} &
  \textbf{Handcrafted Features} &
  0,081 &
  0,023 &
  \textbf{0,948} &
  0,077 &
  0,070 &
  0,927 \\ \cline{2-8} 
\multirow{-3}{*}{\cellcolor[HTML]{FFFFFF}\textbf{Brick Barefeet}} &
  \textbf{VGGish embeddings + hancrafted features} &
  0,069 &
  0,017 &
  \textbf{0,957} &
  0,065 &
  0,067 &
  0,934 \\ \hline
\cellcolor[HTML]{FFFFFF} &
  \textbf{VGGish embeddings} &
  0,645 &
  0,333 &
  0,511 &
  0,133 &
  0,088 &
  \textbf{0,889} \\ \cline{2-8} 
\cellcolor[HTML]{FFFFFF} &
  \textbf{Handcrafted Features} &
  0,048 &
  0,022 &
  \textbf{0,965} &
  0,060 &
  0,059 &
  0,940 \\ \cline{2-8} 
\multirow{-3}{*}{\cellcolor[HTML]{FFFFFF}\textbf{Brick Shoe}} &
  \textbf{VGGish embeddings + hancrafted features} &
  0,032 &
  0,022 &
  \textbf{0,973} &
  0,073 &
  0,055 &
  0,936 \\ \hline
\cellcolor[HTML]{FFFFFF} &
  \textbf{VGGish embeddings} &
  0,640 &
  0,316 &
  0,522 &
  0,052 &
  0,113 &
  \textbf{0,917} \\ \cline{2-8} 
\cellcolor[HTML]{FFFFFF} &
  \textbf{Handcrafted Features} &
  0,085 &
  0,026 &
  \textbf{0,945} &
  0,071 &
  0,064 &
  0,932 \\ \cline{2-8} 
\multirow{-3}{*}{\cellcolor[HTML]{FFFFFF}\textbf{Brick Slipper}} &
  \textbf{VGGish embeddings + hancrafted features} &
  0,056 &
  0,016 &
  \textbf{0,964} &
  0,048 &
  0,073 &
  0,940 \\ \hline
\cellcolor[HTML]{FFFFFF} &
  \textbf{VGGish embeddings} &
  0,716 &
  0,202 &
  0,541 &
  0,132 &
  0,147 &
  \textbf{0,861} \\ \cline{2-8} 
\cellcolor[HTML]{FFFFFF} &
  \textbf{Handcrafted Features} &
  0,096 &
  0,034 &
  \textbf{0,935} &
  0,089 &
  0,101 &
  0,905 \\ \cline{2-8} 
\multirow{-3}{*}{\cellcolor[HTML]{FFFFFF}\textbf{Carpet Barefeet}} &
  \textbf{VGGish embeddings + hancrafted features} &
  0,052 &
  0,036 &
  \textbf{0,956} &
  0,085 &
  0,097 &
  0,909 \\ \hline
\cellcolor[HTML]{FFFFFF} &
  \textbf{VGGish embeddings} &
  0,589 &
  0,378 &
  0,517 &
  0,160 &
  0,126 &
  \textbf{0,857} \\ \cline{2-8} 
\cellcolor[HTML]{FFFFFF} &
  \textbf{Handcrafted Features} &
  0,116 &
  0,051 &
  \textbf{0,917} &
  0,104 &
  0,085 &
  0,906 \\ \cline{2-8} 
\multirow{-3}{*}{\cellcolor[HTML]{FFFFFF}\textbf{Carpet Shoe}} &
  \textbf{VGGish embeddings + hancrafted features} &
  0,056 &
  0,048 &
  \textbf{0,948} &
  0,093 &
  0,086 &
  0,910 \\ \hline
\cellcolor[HTML]{FFFFFF} &
  \textbf{VGGish embeddings} &
  0,548 &
  0,404 &
  0,524 &
  0,129 &
  0,141 &
  \textbf{0,865} \\ \cline{2-8} 
\cellcolor[HTML]{FFFFFF} &
  \textbf{Handcrafted Features} &
  0,069 &
  0,043 &
  \textbf{0,944} &
  0,115 &
  0,107 &
  0,889 \\ \cline{2-8} 
\multirow{-3}{*}{\cellcolor[HTML]{FFFFFF}\textbf{Carpet Slipper}} &
  \textbf{VGGish embeddings + hancrafted features} &
  0,050 &
  0,038 &
  \textbf{0,956} &
  0,130 &
  0,104 &
  0,883 \\ \hline
\end{tabular}

\label{tab:transfer-learning}
\end{table*}

\subsection{Sensor Multiplexing}
Unlike other wearable devices, earbuds possess a unique advantage of being able to sense with both the left and right earbud simultaneously, providing a sensor multiplexing gain. Taking the data collected when subjects walked on tiles with sneakers as an example, we study the performance of using solely the left or right earbud, as well as the achievable gain with both earbuds. As presented in~\Cref{fig:left_right_impact}. when a single earbud is used, the left and right one achieves quite similar performance on the three metrics, suggesting both of them are effective in detecting user' gait. When combining the features extracted from both earbuds, the BAC lifts from 90.5\% to 96.8\%, indicating a great multiplexing gain when using two earbuds.    

\subsection{Impact of training size}
\label{sec:training_size}
To assess the overhead of \SysName during enrollment phase, we study the impact of training data size on the \textcolor{black}{identification} performance as the minimal number of gait cycles required for training is the actual burden on users. Specifically, we select 80 gait cycles for each subject from the tiles-sneakers dataset and run the experiment with Bi-SVM (Balanced-5) protocol. 20\% (16) gait cycles are fixed as the testing dataset to ensure a fair comparison. Then, we continuously increase the number of training samples from 10\% (8) to 80\% (64). The experiment runs when one subject is iteratively selected as the legitimate user and \Cref{fig:training_size} shows the averaged performance over the 31 subjects. Obviously, the performance improves with the increase of training samples. The BAC reaches 93.5\% when 32 gait cycles are used for training. Based on the walking speed of participants, the corresponding overhead for data collection is around 30~s continuous walk, which would be acceptable for most people. 

In addition, \SysName can adapt its \textcolor{black}{identification} model by re-training it with new gait samples. For example, if a new gait cycle is recognized as positive/negative with high confidence, it can be added to the training set for model re-training. Thus, the performance of \SysName is expected to improve continuously after it has been put into practical use.

\subsection{Transfer Learning}
\label{sec:transfer-learning}
We sought to improve the performance of EarGate by looking at more complex, deep Learning based approaches.
Although the performance of traditional deep learning techniques are hindered by the modest size of our dataset (the number of available data points is not sufficient for the model to properly learn the weights\footnote{Considering VGGish as an example, the number of trainable parameters is 4,499,712.}), transfer learning may come handy~\cite{pan2009survey}.
In doing so, we leveraged VGGish, an audio-specific pre-trained model released by Google~\cite{hershey2017cnn}.
VGGish is a convolutional neural network (CNN) that is trained using a large-scale YouTube audio dataset.
Concretely, we use VGGish to automatically extract audio features (\textit{embeddings}) from the raw in-ear audio data.
At this point, we combined the handcrafted features with the embeddings extracted by VGGish and trained two different classifiers: SVM (as in our previously reported results) and XGBoost, an advanced decision-tree based machine learning algorithm~\cite{chen2016xgboost}.
We report our findings in~\Cref{tab:transfer-learning}.
As we can appreciate from ~\Cref{tab:transfer-learning}, XGBoost consistently outperforms SVM when solely using the VGGish embeddings, without any handcrafted feature.
Conversely, by feeding handcrafted features to SVM we achieve better results (both in terms of BAC as well as, in most cases, in terms of FAR and FRR).
Notably, by combining the embeddings extracted with transfer learning and the manually engineered features, we can further boost the performance of EarGate, increasing the BAC and lowering both FAR and FRR.
This is due to the model learning from both the carefully selected features as well as from the more abstract representation of the data generated by VGGish.
Interestingly, while in terms of accuracy the benefit of transfer learning is clear, it is also important to bear in mind the accuracy versus power consumption trade-off.
The performance achievable with only the handcrafted features are indeed only marginally lower than those achieved by combining the transfer learning embeddings and the manually crafted features, in face of an inevitably higher power consumption (caused by the execution of the VGGish model).
However, if a larger dataset is available, the performances associated with transfer learning could further improve.

\subsection{Contribution of Specific Features}
\label{sec:features-contribution}
Our classifier was trained on a variety of features, including many related to the frequency spectrum. 
For an evaluation of our scale, it is likely that walking cadence will be distinct for all participants and, therefore, there is the risk that the model learned to strongly weight the frequency features closely associated with people's cadence. 
This would be problematic because walking cadence will not be distinct on the broader population level, and furthermore people move at different cadences (e.g., when walking alongside someone). 

To confirm that our classifier was not simply a cadence classifier, we asked one subject to walk at three \textit{speeds}: slow (1.59 step/s), normal (1.97 step/s), fast (2.13 step/s). 
After having trained the model with normal-speed walking data, we tested it on all the three instances. 
\Cref{fig:speed_feature} reports the BAC for normal, slow, and fast pace respectively: 99.25\%, 98.30\%, and 96.97\% BAC. These results clearly show that the model is not biasing towards cadence as the identifier, and is instead learning something more fundamental to the user's movement.

Given this we also considered the value of the different features. After training our classifier individually on each of the features reported in~\Cref{sec:features}, we plot in~\Cref{fig:speed_feature} the BAC achieved training our system with only Mel-Frequency Cepstral Coefficients (MFCC) as well as with all the features. Notably, among all the features taken individually, MFCC achieved the best results.
Interestingly, when analyzing the computation time of different features, we found that most of the feature extraction time (89\%) is actually consumed on a feature called 'tonnetz' (with 6 values). So, we repeated the experiment after removing this feature and the identification accuracy almost remains the same.
Therefore, it is possible to use a smaller set of features and reduce the overall end-to-end latency of the system.

\begin{table*}[t]
\small
\centering
\caption{Power consumption and latency measurement of EarGate.}
\vspace{-0.15in}
\label{tab:power}
\setlength\tabcolsep{6.0pt}
\begin{tabular}{ccccc}
\toprule

\textbf{Scheme}& \textbf{Operation} & \textbf{Power (mW)} &  \textbf{Latency (ms)} & \textbf{Energy (mJ)}\\ \cmidrule{1-5}
 
 & MicRecd & 120 & 1000 &   \\
On-device & LowPassFilt & 635 & 1.83 & 168.59 (All)  \\
\textcolor{black}{identification} & FeatExtr (All/MFCC) & 655/651 & 71.98/23.62 & 136.82 (MFCC)   \\
 & Inference & 644 & 0.44 &   \\ \cmidrule{1-5}

& MicRecd & 120 & 1000 &   \\
Raw Data& TX [OS+Air] (WiFi) & 334 & 9.49+12.8 &  123.17 (WiFi)\\
Offloading& TX [OS+Air] (BT) & 478 & 148.41+128 &  190.94 (BT)\\
\cmidrule{1-5}

& MicRecd & 120 & 1000 &   \\
& LowPassFilt & 635 & 1.83 & 168.91 (WiFi, All) \\
Feature& FeatExtr (All/MFCC) & 655/651 & 71.98/23.62 & 172.27 (BT, All) \\
Offloading& TX [OS+Air] (WiFi)(All/MFCC) & 332 & 1.81+0.59/0.39+0.26 & 136.67 (WiFi, MFCC) \\
& TX [OS+Air] (BT)(All/MFCC) & 457 & 8.66+5.94/5.74+2.56 & 139.26 (BT, MFCC) \\
    
\bottomrule 
\end{tabular}
\end{table*}

\section{System Considerations}\label{sec:system_eval}
To gauge the system-level performance of EarGate, we conducted a power-consumption and latency investigation using the same prototype described in~\Cref{sec:implementation}. We assume the model is pre-trained and only consider the real-time \textcolor{black}{identification} overhead.
Given \SysName can either run the \textcolor{black}{identification} framework on-device, or offload it, we consider three different schemes which would impact differently power consumption and latency:
\begin{enumerate}
    \item \textbf{On-device \textcolor{black}{identification}}: microphone data recording (MicRecd), as well as all the gait \textcolor{black}{identification} procedures, including low-pass filtering (LowPassFilt), feature extraction (FeatExtr), and inference, are performed on-device.
    \item \textbf{Distant \textcolor{black}{identification} (raw data offloading)}: the earable records microphone data and directly transmit the raw data via WiFi or Bluetooth (BT), without any processing. The size of raw data is 16~KB~\footnote{With a sampling rate of 4000~Hz and gait cycle length of 1 second, the data from two microphones is $2\times4000\times2B=16~KB$. }. The \textcolor{black}{identification} process would be performed on the cloud or companion smartphones and possibly the result will be communicated back to the device. 
    \item \textbf{Distant \textcolor{black}{identification} (features offloading)}:  both low pass filtering and feature extraction are carried out on the earable, right after recording the microphone data. Only the extracted features (either all features or only MFCC features) are transmitted via WiFi/BT and the result might be communicated back to the device. The size of features is 0.748~KB for all features and 0.16~KB for MFCC features~\footnote{All features includes 187 feature from each microphone data and the size is $2\times187\times2B=0.748~KB$, MFCC features includes 40 features and the size is $2\times40\times2B=0.16~KB$. }.
\end{enumerate}

Our aim here is to show the flexibility of our system to work, irrespective of the network architecture preferred or available. 

For the offloading cases, we consider two types of radio frequency (RF) communications: Bluetooth (BT) and WiFi, as they are (or will soon be) the commonly available radio chips in earables.
For WiFi, we consider a typical uplink throughput of 10~Mbps (similar to that of a domestic network) to compute the transmission latency over the air. 
Regarding Bluetooth, the version supported by the Raspberry Pi we use is BT 4.1. 
Compared to more recent BT standards (e.g., Bluetooth 5, available in Apple AirPods Pro), BT 4.1 offers less throughput (1~Mbps instead of 2~Mbps). As a consequence of that, the over-the-air transmission time reported is longer than what it would be if a more advanced version of BT were adopted. 
We measure the power consumption with a USB power meter. 
Latency measurements are obtained timing the execution of the software handling the operation of interest. 
The results are averaged over multiple measurements and presented in~\Cref{tab:power}. 
The baseline power consumption of our Raspberry Pi (idle) is around 2,325~mW and the values reported in the table are additional power consumption. 
The energy column computes the total energy required to perform the operations for one gait cycle. 

On-device \textcolor{black}{identification} performs the whole pipeline including microphone recording (MicRecd), filtering (LowPassFilt), feature extraction (FeatExtr), and inference on the device. 
The power column indicates that numerical computations (LowPassFilt, FeatExtr, and Inference) are intensive and more power-hungry than microphone recording. 
The latency column shows that most of the processing time is spent on feature extraction. 
Regardless of the time for data acquisition, the overall \textcolor{black}{identification} latency is within 100~ms. 
Concretely, the energy required for a one-time on-device \textcolor{black}{identification} is 168.58~mJ (using all features) and 136.82~mJ (using MFCC only, the most effective feature as described in \Cref{sec:features-contribution}).

When offloading the raw data to the cloud, only MicRecd is performed on-device.
Here, we consider the latency as the sum of TX OS (the time that Pi requires to write the data in the buffer of the chosen interface, either Bluetooth or WiFi) and TX Air (the time it takes for the data to propagate over-the-air). 
We can observe that the latency is largely dependent on the throughput of the network. 
The energy required for WiFi offloading is 123.17~mJ, whilst for BT is 190.94~mJ.
Conversely, when offloading the features, also LowPassFilt and FeatExtr are done on-device. 
Here, following~\Cref{sec:features-contribution}, we compare the energy efficiency of sending all the features or just the MFCC features. 
Given there are 187 features in total and only 40 are MFCC features, the transmission latency for MFCC features only is shorter, overall below 100~ms. 

In summary, we can conclude that (1) both on-device \textcolor{black}{identification} and features offloading guarantee a time delay (after data acquisition) of less than 100~ms, showing EarGate can work in real-time scenarios both on-device and while offloading features;
(2) with respect to the energy consumption, all the schemes consume less than 200~mJ for one-time \textcolor{black}{identification};
(3) the latency and energy consumption for raw data offloading are strictly dependent on the quality of the communication link. 
Thus, offloading raw data would be the best option when the network is stable.
Notably, in this section we focus on the SVM-based pipeline, only considering the case where handcrafted features are used to train the model.
The reasons we do that are manifold.
First, at the moment, deploying a complex model like VGGish on resource-constrained devices (like earbuds) is an extremely challenging task and there are no publicly available tools to supporting it.
Second, given the performance improvements of the transfer learning approach over the handcrafted features one are marginal (\Cref{sec:transfer-learning}).
Third, running VGGish to extract the embeddings would entail far more operations than simply leveraging a limited set of handcrafted features to train SVM and, therefore, it is fair to assume that power and latency figures would be considerably higher than those for the traditional machine learning pipeline.
\section{Discussion}\label{sec:discussion}
In this section, we talk through the limitations of our current work, the possible improvements, and the future directions we plan on pursuing.
Despite the promising results, we are aware of some of the limitations of our approach.
First and foremost, in order for our system to work, initial user data (in the enrollment phase) are required. 
While we acknowledge it would be ideal if such a phase did not have to take place, most of the well-established biometric \textcolor{black}{identification} schemes, like facial recognition and finger-print, do require some initial user data.
Besides, in our evaluation (\Cref{fig:training_size}) we show we only need a limited amount of user data to start offering acceptable performance. 

Second, whilst~\Cref{fig:shoe_ground_impact} shows the impact of different footwear is marginal, when the model is trained on them, to maintain high performance, the model should be partially re-trained to add new pairs of shoes to the legitimate user data. This could be done by the user walking and manually committing the new gait cycles to re-train the model; or it could happen in background if the user changes shoes while wearing the earable (provided the user has been \textcolor{black}{successfully identified} by the earable).
We believe the latter is a reasonable assumption, especially for hearing aid users as such devices are continuously worn throughout the day. Hopefully, this will soon be the case for leisure earables, too, which, once the advancements in materials will guarantee better comfort, could also be worn for a longer amount of time. 
Further, although gait may slightly change over the years, continuously adapting the model (like we do for different shoes) could relieve this issue, too.
Alternatively, the impact of different footwear could be obviated by mean of the combination of a general model and a personalized model.
For instance, an auto-encoder~\cite{li2015hierarchical} could be trained on all the users' data to obtain a general model.

Lastly, one other concern could be related to the obstruction of the ear canal orifice, and the consequent impact on audible sounds (which will result muffled due to the presence of an obstructing body).
We are aware this could be a potential safety issue, therefore, similarly to the AirPods Pro \textit{Transparency Mode}, it is possible to do the same with EarGate.
By playing back the audio recorded by the external mic, the use will be able to hear as if there were no earbuds obstructing their ear canal.
Notably, this does not affect our system as we can easily filter it out (leveraging the difference in frequency), like we did for music. In addition, we believe the power consumption and latency can be further reduced when specialized audio chips (e.g., Apple H1 Chip) are used and more advanced engineering work (e.g., dedicated PCB design) is implemented.
\section{Related Work}\label{sec:related_work}
Gait is a well-studied human biometric, proven to be unique for each individual, and, therefore, often used as \textcolor{black}{an identification} biometric~\cite{wan2018survey, marsico2019survey}.
Traditional approaches for gait recognition enumerate machine vision-based approaches~\cite{zhao20063d}, floor sensor-based techniques~\cite{middleton2005floor}, wireless fingerprinting based method~\cite{wang2016gait}, and wearable sensor-based methods~\cite{gafurov2006biometric,ma2020simultaneous}. These approaches own specific advantages (e.g., zero user effort and complete device-free) and disadvantages (high computation cost and privacy issue for vision-based method, and requirement of deploying the wireless transceivers for wireless fingerprinting based method), thereby complementing each other in different scenarios.

Different from the more traditional techniques, there are two works using acoustic as the modality for gait recognition. Geiger et al.~\cite{geiger2014acoustic} exploited a microphone attached to the foot to record the walking sounds when human feet hit the ground. The main drawback is that the step sounds change with different materials of the ground or shoe sole. In the extreme case, the microphone may not be able to observe noticeable sound when walking on the carpet with barefoot. Instead, our work leverages the occlusion effect to measure the bone-conducted sounds (essentially vibrations) in the ear canal, which are robust to footwear and ground material. 
Wang et al.~\cite{wang2018gait} proposed a fingerprinting-based system (called Acoustic-ID) for human gait detection using acoustic signals. Specifically, by deploying a pair of acoustic transmitter (ultrasound) and receiver, gait pattern is extracted by measuring the reflected acoustic variations when human is walking within the sensing range. Unlike Acoustic-ID that requires to actively transmit ultrasound, our approach is completely passive and does not require the deployment of any additional hardware.

Instead of using gait, researchers have discovered other biometric acquired from human ear for person identification. Nakamura et al.~\cite{nakamura2017ear} proposed to \textcolor{black}{identify and} authenticate users with in-ear electroencephalogram (EEG) measured by a customized earpiece. More recently,  EarEcho~\cite{gao2019earecho} leverages the uniqueness of ear canal geometry to \textcolor{black}{recognize} the users.
Our work differs from EarEcho in three aspects. First, the operation rationale is different. EarEcho is based on the uniqueness of the geometry of the ear canal, while EarGate leverages the uniqueness of the human gait to identify the user. Second, EarEcho requires the active transmission of a sound/ultrasound pulses so as to measure the echo sound, while EarGate is completely passive. Third, as demonstrated in~\cite{amesaka2019facial}, the geometry of the ear canal will change under different facial expressions, which might impact the effectiveness of EarEcho in daily life. Although gait can also change over time, it usually evolves in a large time span, so that the user can periodically adapt the trained model.
Moreover, the uniqueness of ear canal geometry has not been tested with a large population (only 20 subjects involved in~\cite{gao2019earecho}).  
Ultimately, with this work, we propose the first earable-based acoustic-gait \textcolor{black}{identification} system, showing its potential applicability to various \textcolor{black}{identification} use cases.
\section{Conclusion}\label{sec:conclusion}
We presented EarGate, an earable \textcolor{black}{identification} system based on user gait. Exploiting the occlusion effect, EarGate enables detection of human acoustic gait from an in-ear facing microphone. Experimenting with 31 subjects, we demonstrated that EarGate achieves robust and acceptable \textcolor{black}{identification} performance (up to 97.26\% BAC, with low FAR and FRR of 3.23\% and 2.25\% respectively) under various practical conditions. Moreover, EarGate will not affect the general functionality of earbuds and is robust to high-frequency noises like music playback and human speech. We envision that EarGate can be an effective and robust way of \textcolor{black}{identifying} (standalone or companion with smartphone) earables. Particularly, its unobtrusiveness makes it an appealing replacement of FaceID for users wearing face masks during the COVID-19 pandemic. 
\section{Acknowledgments}

This work is supported by ERC through Project 833296 (EAR) and by Nokia Bell Labs through a donation. We thank D. Spathis and L. Qendro for the insightful discussions, the anonymous shepherd and reviewers for the valuable comments, and the volunteers for the data collection.

\balance
\bibliographystyle{ACM-Reference-Format}
\bibliography{references}

\end{document}